\documentclass[reqno,11pt,twoside]{article}  
\newcommand{\version}{January 17, 2012}
\renewcommand{\mark}[1]{}
\newlength{\dinwidth}
\newlength{\dinmargin}
\usepackage{amsmath}
\usepackage{amssymb}
\usepackage{latexsym}
\usepackage{cite}
\usepackage{epsfig,psfrag}
\newcommand{\RR}{\mathbb{R}}
\newcommand{\CC}{\mathbb{C}}
\newcommand{\II}{\mathbb{Z}}

\newcommand{\calA}{{\mathcal A}}
\newcommand{\calF}{{\mathcal F}}

\newcommand{\calH}{{\mathcal H}}
\newcommand{\calO}{{\mathcal O}}

\newcommand{\calS}{{\mathcal S}}

\newcommand{\half}{{\frac{1}{2}}} 
\renewcommand{\d}{d}
\newcommand{\unity}{1}
\newcommand{\bfp}{{\boldsymbol{p}}}

\newcommand{\clo}{ {\mbox{\bf --}} }
 
\newcommand{\spa}{{\rm span }} 

\newcommand{\supp}{{\rm supp}}
\newcommand{\eps}{\varepsilon}

\newcommand{\lsp}{(\,}
\newcommand{\rsp}{\,)}
\newcommand{\Lsp}{\big( \,}
\newcommand{\Rsp}{\,\big)}
\newcommand{\FA}{\calF}     

\newcommand{\spd}{e}   
\newcommand{\spdpath}{\tilde{e}}   
\newcommand{\Spd}{H}   
\newcommand{\spc}{C}   
\newcommand{\cckpath}{\tilde{K}}    
\newcommand{\cck}{K}    
\newcommand{\spcpath}{\tilde{\spc}}    

\newcommand{\V}{\Gamma}     

\newcommand{\In}{_{\text{\rm in}}}      

\newcommand{\timesIn}{\stackrel{\text{\rm in}}{\times}} 
\newcommand{\timesOut}{\stackrel{\text{\rm out}}{\times}}

\newenvironment{Proof}%
{\par \medskip \noindent {\em Proof.}}{\hspace*{\fill} $\square$\par%
\medskip\noindent}
\newtheorem{Thm}{Theorem}
\newtheorem{Prop}[Thm]{Proposition} 
\newtheorem{Lem}[Thm]{Lemma} 
\newtheorem{Cor}[Thm]{Corollary} 

\newcommand{\beq}{\begin{equation}} 
\newcommand{\eeq}{\end{equation}} 
\newcommand{\Eps}[2]{\eps_{#1 #2}} 
\newcommand{\tamp}{\check t_{\rm amp}}

\newcommand{\ramp}{\check r_{\rm amp}}
\newcommand{\wamp}{\check w_{\rm amp}}

\newcommand{\aamp}{\check a_{\rm amp}}
\newcommand{\Hyp}{H_m^+}
\newcommand{\Hypp}{H_m^2}
\newcommand{\pibar}{\bar \pi}

\begin{document}
\title
{Braid group statistics implies scattering \\ 
in three-dimensional local quantum physics} 
\author{Jacques Bros$^1$ and Jens Mund$^2$\thanks{Supported by CNPq.}\\ 
\\
$^1$Institut de Physique Th\'eorique, CEA -- Saclay, 
France  \\
$^2$Departamento de F\'{\i}sica, UFJF, 
Juiz de Fora, Brazil
} 
\date{ \version \\ \vspace{3ex} 
}
\maketitle 
\begin{abstract}
It is shown that particles with braid group statistics (Plektons) in
three-dimensional space-time cannot be free, in a quite 
elementary sense: They must exhibit elastic 
two-particle scattering into every solid angle, and at every energy. 
This also implies that for such particles there cannot be any operators 
localized in wedge regions which create only
single particle states from the vacuum and which are well-behaved
under the space-time translations (so-called temperate 
polarization-free generators). 
These results considerably strengthen an earlier ``NoGo-theorem for
'free' relativistic Anyons''.

As a by-product we extend a fact which is well-known in quantum field
theory to the case of topological charges (i.e., charges localized 
in space-like cones) in $d\geq 4$, namely: 
If there is no elastic two-particle 
scattering into some arbitrarily small open solid angle element, then the
2-particle S-matrix is trivial. 
\end{abstract}
\maketitle 
\section{Introduction} 
\label{secIntro} 
In relativistic quantum field theory in $d>2$ space-time dimensions 
it is a well-established fact~\cite{Araki,BEG64} that 
if there is some arbitrarily small open solid angle element into which 
there is no elastic two-particle scattering, then 
the $2\to 2$ particle S-matrix is trivial, and then there is also no
particle production. We show that the same holds in the quite general 
setting of local quantum physics admitting topological charges in
$d\geq 4$. 
Our main result, however, is that the hypothesis {\em cannot be
  satisfied} for Plektons in $d=3$, 
namely: Braid group statistics implies elastic two-particle scattering. 
This also implies that for Plektons there cannot be a model which has  
temperate polarization-free generators in the sense of Borchers et
al.~\cite{BBS}. (These are operators localized in wedge regions which 
create only single particle states from the vacuum and which are well-behaved
under the space-time translations.)
These results considerably strengthen an earlier ``NoGo-theorem for
'free' relativistic Anyons'' by the one of the authors~\cite{M}.

In view of our finding that there are no free Plektons, the 
Lagrangean approach does not seem appropriate for the construction 
of relativistic quantum fields with braid group statistics, since 
its underlying strategy 
is the coupling of free fields to other fields or among themselves. 
In fact, the situation with respect to Lagrangean relativistic model
building is quite unsatisfactory. 
Lagrangean local gauge theories have been proposed (for more than 20 
years) where the matter fields assume anyonic properties by coupling 
to a Chern-Simons type gauge field~\cite{Semenoff,Swanson,Jack,Ban,Mintchev91}. 
However, except for lattice models~\cite{BaNi95,Lusch,Mull}, these
models have not been made completely explicit. Consequently, as 
S. Forte put it~\cite[p.~235]{Forte}, 
different and contradictory conclusions have been drawn from the 
same starting point (Lagrangean). 

We consider the scattering of two massive charged
particles, 
assuming that the respective masses are isolated values in the 
mass spectra within the respective charge sectors.
We admit that the particles carry topological 
charges,  
that is, charges which are localizable not in compact regions but only
in space-like cones\footnote{A space-like
  cone is a region in Minkowski space of the 
form $\spc=a+\cup_{\lambda>0}\lambda \calO,$
  where 
  $a\in\RR^4$ is the apex of $\spc$ and $\calO$ is a double cone whose
  closure is causally separate from the origin.}, which is the worst possible
dis-locality in the purely massive case~\cite{BuF}. 
For simplicity, we assume that the masses of the two particles coincide. 
Given two single particle states $\psi_1$, $\psi_2$ in respective 
charge sectors $\pi_1$, $\pi_2$, and having disjoint 
energy-momentum supports, Haag-Ruelle scattering theory associates an 
outgoing and an incoming $2$-particle scattering state 
$$
\psi_1\timesOut \psi_2, \quad \psi_1\timesIn \psi_2. 
$$ 
In the case of permutation group statistics, these scattering states depend
only on the single particle states $\psi_k$, as indicated by the
notation. However, in the case of braid group statistics they also
depend on the space-time localization regions in which the charged  
states $\psi_k$ have been ``created from the vacuum''.  
Now suppose that there are sets of incoming momenta 
$U_1,U_2$ and of outgoing momenta $V_1,V_2$ on the mass shell, 
mutually disjoint: 
\beq \label{eqMutualDisjoint}
U_1\cap U_2=V_1\cap V_2=U_i\cap V_j=\emptyset,
\eeq 
for which all scattering amplitudes in the channel
$\pi_1\times \pi_2\to \pi_1\times \pi_2$ are trivial --- even though the
corresponding process is admitted by energy-momentum conservation,
namely   
\beq \label{eqUV}
(U_1+U_2) \cap (V_1+V_2) \neq \emptyset.  
\eeq
More precisely, for all single-particle states $\phi_i$ in the sectors
$\pi_i$ with respective spectral supports in $U_i$ and $\psi_i$
in the same sectors $\pi_i$ and with supports in $V_i$ 
and for some admissible space-time localization regions\footnote{Only
  relevant in the case of braid group statistics.} there holds 
\begin{equation} \label{eqStriv}
\Lsp \phi_1\timesIn \phi_2\,,\, \psi_1\timesOut \psi_2 \Rsp
= 0. 
\end{equation} 
We then say that there is {\em no elastic two-particle scattering from
$U_1\times U_2$ into $V_1\times V_2$} in the channel $\pi_1\times
\pi_2\to \pi_1\times \pi_2$.\footnote{Contrapositively, we say that 
there {\em is}  elastic two-particle scattering 
from $U_1\times U_2$ into $V_1\times V_2$  
if  for every admissible space-time localization
region there are some $\phi_i,\psi_i$ such that the above scattering 
amplitude is non-zero.} 

Our first result is that, in space-time dimension $d\geq 4$, 
this hypothesis implies triviality of the two-particle S-matrix.  
For simplicity of the argument, we suppose that a pair of disjoint  
sets of incoming and outgoing momenta for which there is 
no scattering exists for every total momentum. 
\begin{Thm}[Triviality of the 2-particle S-matrix in $d\geq 4$.]  
\label{STrivial} 
Suppose that for every total momentum $P$, $P^2\geq 4m^2$, there
are non-empty open subsets $U_1,U_2,V_1,V_2$ of the mass shell, mutually
disjoint, such that $P\in\, (U_1+U_2) \cap (V_1+V_2)$,
and such that there is no elastic two-particle scattering from 
$U_1\times U_2$ into $V_1\times V_2$ in the channel 
$\pi_1\times \pi_2\to \pi_1\times \pi_2$. 
Then the $2$-particle S-matrix in this channel is trivial, i.e.~for
every pair of single particle states $\psi_1,\psi_2$ in the 
respective 
sectors $\pi_1,\pi_2$ the incoming and outgoing scat\-tering 
states coincide: 
\beq \label{eqIn=Out} 
\psi_1\timesIn \psi_2 = \psi_1\timesOut \psi_2. 
\eeq
\end{Thm}
This extends a well-established fact from the case of 
compactly localized charges to the case of topological charges. 
We consider this result of interest in its own right in view of 
the various recent constructions of "non-local" 
quantum field 
theories~\cite{GrosseLechner08,Lechner11,BuSu_Warped}.  
{}For example, the 
deformation approach constructs theories with non-trivial 
S-matrices satisfying the hypothesis. Our theorem then gives a
structural argument that in these models the 
spacelike-cone-algebras are too small as to satisfy the 
Reeh-Schlieder property --- which otherwise has to be (and of 
course has been) checked by explicit calculations. 
The theorem is also needed in a recent algebraic version of the 
Jost-Schroer theorem~\cite{Mu_JoSch}. 

Our main result concerns models with topological charges in
three-dimensional space-time, where braid group statistics may occur. 
Here we consider the case where $\pi_1$ and $\pi_2$ are conjugate 
charges or, in the Abelian case, coincide. 
It turns out that the hypothesis~\eqref{eqStriv} implies that
the statistics parameter of the corresponding sector is real, 
thereby excluding braid group statistics. In other words, we show:  
\begin{Thm}[Braid group statistics implies scattering.]  \label{WW}
Let $\pi$ be a  massive single particle representation 
with braid group statistics\footnote{That is to say, with non-real
  statistics parameter, see Sec.~\ref{secSetting}
}, and let $\bar \pi$ be the 
conjugate sector. Then for any non-empty open sets of energy-momenta 
$U_1,U_2,V_1,V_2$ admitted by energy-momentum
conservation~\eqref{eqUV} there is elastic 
two-particle scattering from $U_1\times U_2$ into $V_1\times V_2$ in 
the channel $\pi\times \bar \pi\to \pi\times \bar\pi$. 
In the case of Anyons, the same holds for the channel $\pi\times
\pi\to \pi \times \pi$. 
\end{Thm}
Now Borchers et al.~\cite{BBS} show that the existence of temperate 
wedge-localized polarization-free generators (PFG's) implies that 
there are
non-empty subsets $U_i$, $V_i$ of the mass shell for which there is no
elastic two-particle scattering even though admitted by 
energy-momentum conservation, see Eq.~(3.27) in~\cite{BBS}, which can be derived also 
in the braid group statistics case. 
We therefore conclude from the above theorem:   
\begin{Cor}[Non-existence of PFG's for Plektons.] \label{PFG} 
The existence of temperate wedge-localized polarization free
generators excludes braid group statistics.  
\end{Cor}
We shall prove these theorems along the following standard lines. 
Due to the assumed disjointness of the energy-momentum
supports, the state $\psi_1\timesIn\psi_2$ also has vanishing
scalar product with $\phi_1\timesIn \phi_2$, and Eq.~\eqref{eqStriv} can be
rephrased by saying that the corresponding S-matrix element is
trivial, 
\beq\label{eqStriv''} 
\lsp\phi_1\timesIn\phi_2\,,\,\psi_1\timesIn\psi_2-\psi_1\timesOut\psi_2\rsp = 0.
\eeq
The first step is now to establish  in the present setting the
relevant LSZ relations, which relate the left hand side of the above equation 
with the amputated time-ordered function~\eqref{eqtamp}. 
The second step is to establish on-shell analyticity properties 
of the amputated time-ordered function. 
These analyticity properties  
are weaker~\cite{BrosEpstein94} than in the case of compact
localization but, as it turns out, still sufficient for the
present purpose: To extend Eq.~\eqref{eqStriv''} to other 
particle configurations $\psi_i$ (for fixed $\phi_i$). 
Namely, in $d\geq 4$ Eq.~\eqref{eqStriv''} extends to {\em all} 
$\psi_i$,
which leads to Theorem~\ref{STrivial}. In $d=3$, it extends to such
$\psi_i$ which have common supports with the states $\phi_i$, so that  
the states $\psi_1\timesIn\psi_2$ no longer have necessarily 
vanishing scalar products with $\phi\timesIn \phi_2$. Then a 
careful analysis of
the dependence of the scattering states $\psi_1\timesOut\psi_2$,
$\psi_1\timesIn\psi_2$ on the
space-time localization regions shows that Eq.~\eqref{eqStriv''} is
incompatible with braid group statistics. 
 
The article is organized as follows. 
Section~\ref{secSetting} is devoted to a more detailed account of the
general setting, and of some known facts on the scattering of
Plektons. 
In Section~\ref{secLSZ} we show that certain LSZ
relations hold true in the case of topological charges, eventually
with braid group statistics, under certain hypothesis on the momentum
supports and space-time localization regions.   
In Section~\ref{secAnalyt} we give a detailed account of the
analyticity domain of the relevant amputated time-ordered (or
advanced, or retarded) functions, see also~\cite{BrosEpstein94}.   
Finally, we prove the two theorems in Section~\ref{secProofs}. 
\section{General Setting}  \label{secSetting}
\label{secFrame} 
An analysis of the S-matrix is most conveniently done in terms of
charge carrying local field operators. In four-dimensional space-time, 
these comprise an algebra $\calF$ which contains the observables
$\calA$ as the sub-algebra of invariants under a global gauge
symmetry. However such a frame, the so-called Wick-Wightman-Wigner
scenario, does not exist in the presence of non-Abelian braid group 
statistics in
$d=3$. As a unified minimal framework for both cases (permutation 
and braid group statistics), we shall use the field bundle
formalism~\cite{BuF,FGR,DHRIV}, which is quite rudimentary but
suffices for the present purpose. 
\paragraph{{\it Particles and topological charges.}}
We consider a collection $\Delta$ of superselection sectors, namely, 
representations of the observable algebra. To each $\pi\in\Delta$
there is associated a Hilbert space $\calH_\pi$ describing the
corresponding charged states. Each $\calH_\pi$ carries a unitary
continuous representation $U_\pi$ of the translation group 
$\RR^d$ satisfying the relativistic
spectrum condition (positivity of the energy). (In the proof of
Thm.~\ref{STrivial} we shall also use Lorentz covariance, but only in
order not to burden the hypothesis.)  
The Hilbert space of the vacuum representation $\pi_0$ contains a
unique (up to a factor) translation invariant vector $\Omega$. 
We are particularly interested in so-called {\em massive single 
particle sectors}, describing the states of massive particles 
carrying topological charges. 
These are sectors $\pi$ whose energy momentum spectrum has an isolated mass 
shell as the lower boundary.\footnote{We thus assume implicitely that
  the vacuum representation does not contain massless particles.}  
The charge described by $\pi$ is then localizable with respect to 
the vacuum in space-like cones~\cite{BuF}. Namely, the representation $\pi$ is equivalent to the vacuum
representation when restricted to the causal complement of any
space-like cone. 
Within the set $\Delta$ of sectors one has intrinsic notions of charge 
composition $\pi_1\times  \pi_2$, and of charge conjugation, 
namely, every charge 
$\pi\in\Delta$ has a conjugate charge $\bar\pi\in
\Delta$, characterized by the fact that the composition $\pi\times
\pibar$ 
contains the vacuum representation. If $\pi$ is a massive
single particle representation, then so is $\bar\pi$, and it describes
the corresponding anti-particles which have the same mass~\cite{F81}
and spin~\cite{BuEp,Mu_SpiSta}. 
There is also an intrinsic notion of exchange statistics relating
$\pi\times \sigma$ and $\sigma\times \pi$. 
Namely, for every pair of localized morphisms $\pi,\sigma$ in
$\Delta$ there is a unitary operator 
\beq \label{eqEps}
\Eps{\pi}{\sigma}:\,\calH_{\pi\times \sigma}\to \calH_{\sigma\times
  \pi} , 
\eeq
the so-called statistics operator. 
The family of statistics operators satisfies the braid 
relations~\cite[Eq.~(2.3)]{FRSII} and in four-dimensional space-time
furnishes a representation of the permutation group due to the fact
that here the monodromy operators $(\Eps{\pi}{\pi})^2$   
are trivial.  In three-dimensional space-time, there may be
non-trivial monodromy operators in which case the statistics operators 
furnish a representation of the braid group.   
Then one speaks of Plektons or, if the corresponding
representation of the braid group is Abelian, of Anyons.   
Associated with each irreducible sector $\pi$ is 
the statistics parameter $\lambda_\pi$ and the statistics phase 
$\omega_\pi$, defined by the relations  
\begin{equation} 
\varphi_\pi(\Eps{\pi}{\pi})=\lambda_\pi \unity, \qquad \omega_\pi =
\frac{\lambda_\pi}{|\lambda_\pi|},  
\end{equation}
where $\varphi_\pi$ is a left inverse for $\pi$~\cite{H96}. 
The statistics parameter is non-real precisely in the case of braid
group statistics. 
In the Abelian case, $\Eps{\pi}{\pi}$ is already trivial, namely, 
$\omega_\pi$ times the unit operator. 
\paragraph{{\it Charge carrying fields.}} 
For every $\pi\in\Delta$ there is a Banach space $\calF_\pi$
of  charge carrying field operators, which act on the hermitean 
vector  bundle 
$\calH\doteq \bigcup_{\sigma\in \Delta} \calH_{\sigma}$ 
in such a way that for every $\sigma\in\Delta$ 
\beq  \label{eqFH}
\calF_\pi:\calH_\sigma \to \calH_{\sigma\times \pi} 
\eeq
acts as a space of linear operators. 
Fields are {localizable} to the same extent to which the charges are
localizable which they carry, namely in space-like cones. 
In the presence of braid group statistics in $d=3$, 
the fields can have definite space-like commutation relations only if
they carry some supplementary information in addition to
the localization region. The possibility we choose is
to consider paths in the set of space-like cones.\footnote{Two 
other possibilities are: 
To introduce a reference space-like cone from
which all allowed localization cones have to keep space-like separated 
(this cone playing the role of a ``cut'' in the
context of multi-valued functions)~\cite{BuF}; or a cohomology theory of 
nets of operator algebras as introduced by 
Roberts~\cite{Roberts,Roberts76,Roberts80}.
} 
(The following considerations are only relevant in $d=3$.) 
Let $\Spd$ be the manifold of space-like directions, 
\begin{equation} \label{eqSpd} 
\Spd :=\{ \spd\in\RR^3,\; \spd\cdot \spd=-1\}.  
\end{equation} 
Every space-like cone $\spc$ is of the form 
$\spc= a+\RR^+ \spc^\Spd$, where $a$ is the apex of $\spc$ and   
$\spc^\Spd \doteq (\spc-a)\cap \Spd$ is the set of space-like 
directions contained in $\spc$. 
We now look upon the space-like cone $\spc$ as the union of all 
``strings'' contained in it, $a+\RR^+ e$, $e\in\spc^\Spd$, 
and further identify such string with the 
pair $(a,e)\in \RR^3\times \Spd$, quite in the sense of 
``string-localized quantum fields''~\cite{Mandelstam,MSY2}. 
Then $\spc$ is identified with a subset of
$\RR^3\times \Spd$:   
\begin{align} \label{eqRH}
\spc 
\leftrightarrow \{a\}\times \spc^\Spd,
\end{align}
where $\spc^\Spd$ is a double cone in $\Spd$~\cite{Mu_BiWiAny}. 
The region $\spc^\Spd$ 
is simply connected, whereas 
$\Spd$ itself has fundamental group $\II$.  
Thus the portion of the universal covering space of 
$\RR^3\times \Spd$ over a region $\spc\cong \{a\}\times \spc^\Spd$ 
consists of a countable infinity of copies (``sheets'') 
of $\spc$. 
As usually, we identify the universal covering space of 
$\Spd$ with homotopy classes of
paths in $\Spd$ starting at some fixed reference direction (the 
``base point'' in $\Spd$).  
We shall generically denote a sheet over $\spc$ by
$\spcpath$ and call it a ``path ending at $\spc$''. 
(It is canonically homeomorphic to $\spc$, but contains in
addition the information of a winding number distinguishing it
from the other sheets over $\spc$.) 
These paths serve to label the localization regions of charged 
fields. 
Namely, for each path $\spcpath$ there is a linear subspace 
$\calF_\pi(\spcpath)$ of $\calF_\pi$, called the fields carrying
charge $\pi$ localized in $\spcpath$. 
This family is isotonous in the sense that 
\begin{equation} \label{eqIsotony} 
\calF_\pi(\spcpath_1)\subset \calF_\pi(\spcpath_2) \qquad 
\text{ if } \qquad \spcpath_1\subset \spcpath_2. 
\end{equation}
(We say that $\spcpath_1 \subset \spcpath_2$ if 
$\spc_1\subset \spc_2$  and
the corresponding paths $\spcpath_1$, $\spcpath_2$ differ by a path 
in $\spc_2$.) 
The vacuum $\Omega$ has the Reeh-Schlieder property for the fields,
i.e.\ for any path of space-like cones $\spcpath$ there holds 
\begin{equation} \label{eqReehSchlie}
\big(\calF_\pi(\spcpath)\,\Omega\big)^\clo = \calH_\pi, 
\end{equation}
where the bar denotes the closure. 
\paragraph{{\it (Braid group) statistics.}} 
The statistics operators determine the commutation 
relations of causally separated fields, as follows. 
Let $d\theta$
be the angle one-form in some fixed Lorentz frame, and for
a path $\spcpath$ let $\theta(\spcpath)$ be the set of corresponding 
``accumulated angles'', namely the interval  
\beq \label{eqAngle}
\theta(\spcpath) \doteq \Big\{\int_{\spdpath}d\theta\,:\; \spdpath\in
\spcpath^\Spd \Big\} \, 
\eeq
on the real line. 
Given two paths $\spcpath_1$, $\spcpath_2$ with $\spc_1$ causally
separate from $\spc_2$, define the {\em relative winding number} 
$N(\spcpath_1,\spcpath_2)$ of 
$\spcpath_2$ w.r.t.\ $\spcpath_1$ to be the unique integer $n$ such that 
\begin{equation*} 
\theta(\spcpath_2) + {2\pi n} \, <\, \theta(\spcpath_1) \, < \, 
\theta(\spcpath_2)+ 2\pi (n+1) . 
\end{equation*}
(Note that this number $n$ is independent of
the Lorentz frame in which $d\theta$ is defined, and of the 
choice of reference direction.)  
Then for every $F_1\in\calF_{\pi_1}(\spcpath_1)$ and  
$F_2\in\calF_{\pi_1}(\spcpath_2)$ there holds the commutation
relation
\begin{equation} \label{eqFieldCRN}
F_1\, F_2 = \Eps{\pi_1}{\pi_2}(\spcpath_1,\spcpath_2) \,F_2\, F_1,
\end{equation}
where\footnote{This follows from Prop.~5.9 in~\cite{FGR}, together 
with the fact that $\pi_0(V_{\varrho_2})=1$ and 
$\pi_0\varrho_1(V_{\varrho_2})=\pi_0 \big(\Eps{\varrho_2}{\varrho_1}
\Eps{\varrho_1}{\varrho_2}\big)$, see~\cite[Eq.~(5.2.3)]{FGR}.
}
\beq
  \Eps{\pi_1}{\pi_2}(\spcpath_1,\spcpath_2) = 
\big(\Eps{\pi_1}{\pi_2}\Eps{\pi_2}{\pi_1}\big)^n\;\Eps{\pi_1}{\pi_2}
\;, \quad
n=N(\spcpath_1,\spcpath_2).  \label{eqEpsN}
\eeq
We shall use below only two special cases, namely winding number $0$ and
$-1$. $N(\spcpath_1,\spcpath_2)=0$ means that 
for $\spdpath_i\in\spcpath_i^\Spd$ the path 
$\spdpath_1 \ast \spdpath_2^{-1}$ is homotopic to a path which 
goes ``directly'' from $\spc_2$ to $\spc_1$ in the mathematically
positive sense.
(Figure~1 shows an example of this situation.) In this case  
$\Eps{\pi_1}{\pi_2}(\spcpath_1,\spcpath_2)=\Eps{\pi_1}{\pi_2}$. 
$N(\spcpath_1,\spcpath_2)=-1$ means that the paths 
$\spdpath_1 \ast \spdpath_2^{-1}$ are homotopic to paths which 
go directly from $\spc_2$ to $\spc_1$ in the mathematically
{\em negative} sense (i.e., the roles of $\spcpath_1$ and 
$\spcpath_2$  in Fig.~1 are interchanged), and in this case  
$\Eps{\pi_1}{\pi_2}(\spcpath_1,\spcpath_2)=\Eps{\pi_2}{\pi_1}^{-1}$. 
(Of course, this is a consequence of the case
$N(\spcpath_1,\spcpath_2)=0$ if Eq.~\eqref{eqFieldCRN} is to be 
consistent.) 
\begin{figure}[ht] 
 \label{Fig1}
\psfrag{e0}{$\spd_0$}
\psfrag{C1}{$\spcpath_1^\Spd$}
\psfrag{C2}{$\spcpath_2^\Spd$}
\psfrag{H}{$H$}
\begin{center}
\epsfxsize20ex 
\epsfbox{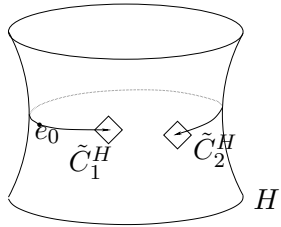}
\caption{The relative winding number $N(\spcpath_1,\spcpath_2)$ 
of $\spcpath_2$ w.r.t.\  
$\spcpath_1$ is zero. ($e_0$ is the base point where the paths 
start.)} 
\end{center}
\end{figure}
\paragraph{{\it Translation covariance.}}
There is an action $x\mapsto \alpha_x$ of the translation group
$\RR^d$ on the fields, under which these are covariant: Namely,
for all $\pi\in\Delta$ and all localization paths 
$\spcpath$ there holds 
\beq \label{eqCov}
\alpha_x:\, \calF_\pi(\spcpath) \to \calF_\pi(x+\spcpath). 
\eeq
(Using the identification~\eqref{eqRH}, the translation $x$ 
does not act on the space-like directions $\spc^\Spd$ contained in
$\spc$ and hence not on $\tilde{\spc}^\Spd$. It only translates the
apex.)  
It is determined by the unitary representations $U_\pi$ acting on 
the Hilbert spaces $\calH_\pi$ via 
\beq
\alpha_x(F) \psi = U_{\sigma\times \pi}(x)FU_\sigma(-x)\psi,\quad 
F\in\calF_\pi,\,\psi\in\calH_\sigma.
\eeq
\paragraph{{\it Scattering states.}} 
Haag-Ruelle scattering theory has been developed 
in~\cite{Hepp65, Jost}, adapted to the 
setting of algebraic quantum field theory in~\cite{DHRIII,BuF}, and to 
theories with braid group statistics in~\cite{FGR}. 
This theory associates to $n$ single particle 
states $\psi_k\in\calH_{\pi_k}$ an outgoing and an incoming 
scattering state 
$$
\psi_1\timesOut \cdots\timesOut \psi_n,\quad \psi_1\timesIn
\cdots\timesIn \psi_n
$$
living in the Hilbert space of the composite sector 
$\pi_1\times  \cdots\times \pi_n$.
In the case of permutation group statistics, these scattering states depend
only on the single particle states $\psi_k$, as indicated by the
notation. However, in the case of braid group statistics they also
depend on the localization regions $\spcpath_k$ in which the charged states
$\psi_k$ have been created from the vacuum. 
Since our argument relies crucially on this fact, we recall the
construction. 

First one constructs quasi-local creation operators as follows. 
Let $\pi\in\Delta$ be a massive single particle sector 
with mass $m$ and let $F\in\FA_\pi(\spcpath)$ be a field
operator which creates from the vacuum a single particle  state with
non-zero probability, i.e., the spectral support of $F\Omega$ has  
non--vanishing intersection with the mass hyperboloid
$H_m^+\doteq\{p:\;p^2=m^2, p_0>0\}$.   
Further, let $f\in\calS(\RR^4)$ be a Schwartz function whose Fourier 
transform\footnote{The Fourier transform of a test 
function $f$ is here $\hat{f}(p)\doteq\int \d^4 x f(x) e^{ip\cdot
  x}$. 
} 
$\hat{f}$ has compact support contained in the open 
forward light cone intersecting the energy momentum spectrum 
only in the mass shell $H_m^+$. 
If this intersection is contained in some $V\subset \Hyp$, namely if 
$$
\supp\hat{f}\cap \text{spec }P_{\pi} \subset \Hyp, \quad  
\supp \hat{f} \cap \Hyp\subset V,
$$
we shall say that $f$ has {\em momentum support} in
$V$. We shall use a fixed reference frame $u\equiv (1,\boldsymbol{0})$
and consider the set of velocities corresponding to $V$ in this
frame, 
\begin{equation}  \label{eqVf}
\V(V) \doteq  \{\, \big(1,\frac{\bfp}{\omega(\bfp)}\big)\,,\;
p=(p_0,\bfp) \in V\,\}\,,    
\end{equation}
where $\omega(\bfp)\doteq\sqrt{\bfp^2+m^2}$. 
For $t\in\RR$, let the function $f_t$ be defined by multiplying the 
Fourier transform with the factor $e^{i(p_0-\omega(\bfp)) t}$.  
For large $|t|$, its support is essentially contained in the 
region $t \,\V(V)$. 
More precisely~\cite{BBS,Hepp65}, for any $\eps>0$ there is a Schwartz function 
${f}^\eps_t$ with support in $t\,\V(V)^\eps,$ where $\V^\eps$
denotes an $\eps$--neighborhood of $\V$, such that
$f_t-{f}^\eps_t$ converges to zero in the Schwartz topology for 
$|t|\rightarrow\infty.$
Consider now the quasi-local operator  
\begin{equation*} 
 F(f_t)\doteq\int \d^4 x \,f_t(x)\, F (x) \,,
\end{equation*}
where $F(x)\doteq \alpha_x(F)$.   
For large $|t|,$ this operator is essentially localized 
in $\spcpath+t\,\V(V).$ 
Namely, for any $\eps>0,$ it can be approximated by the operator 
\begin{equation} \label{eqBeps}
F(f^\eps_t)\in\calF_{\pi}\big(\spcpath+t\,\V(V)^\eps\big) 
\end{equation}
in the sense that  $\|F(f^\eps_t)-F(f_t)\|$ is of fast decrease in $t.$ 
This operator creates from the vacuum a single particle vector 
$F(f_t) \,\Omega= 
\hat{f}(P_\pi)\,F\,\Omega$, independent of $t$. ($P_\pi$ denotes the 
generator of the translations $U_\pi(x)$.) 

Consider now $n$ massive single particle sectors $\pi_k$,
$k=1,\ldots,n$, with mass $m$.\footnote{For notational 
simplicity we consider coinciding masses.} 
To construct an outgoing scattering state from $n$ corresponding single
particle vectors, pick $n$ localization regions  $\spcpath_k$ 
and compact sets $V_k$ on the mass shell, such that the regions 
$\overline{\spc_k}+t\,\Gamma(V_k)$ are mutually space-like separated
for large $t$.\footnote{$\overline{\spc}$ denotes the closure of
  $\spc$. Taking the closure makes sure that for suitable open 
neighborhoods $\V_k^\eps$ of $\Gamma(V_k)$ in $\RR^d$ the regions 
$\spc_k+t\,\V_k^\eps$ are still mutually space-like separated for large
  $t$.} 
We then say that $\spc_1,\ldots,\spc_n$ are {\em future-admissible} 
for $V_1\times\cdots\times V_n$.  
Next, choose $F_k\in\calF_{\pi_k}(\spcpath_k),$ and Schwartz functions
$f_k$ with momentum supports in $V_k.$ Then the standard lemma of
scattering theory, adapted to braid group statistics~\cite{FGR},
asserts the following: The limit 
\begin{equation}  \label{eqBn1Om}
\lim_{t\rightarrow\infty} F_1(f_{1,t})\cdots F_n(f_{n,t})\,\Omega \doteq
(\psi_1,\spcpath_1)\timesOut\cdots\timesOut(\psi_n,\spcpath_n)\, 
\end{equation}
exists and depends only on the single particle vectors 
$\psi_k\doteq F_k(f_{k})\,\Omega$ and, in the case of braid group
statistics, on the localization regions $\spcpath_k$. 
Similarly, we say that $\spc_1,\ldots,\spc_n$ are {\em past-admissible} 
for $V_1\times\cdots\times V_n$ if the regions 
$\overline{\spc_k} + t \Gamma(V_k)$ are mutually causally separated 
for large $|t|$ with $t<0$. In this case, the incoming scattering
state  
\begin{equation}  \label{eqBn1Om'}
\lim_{t\rightarrow-\infty} F_1(f_{1,t})\cdots F_n(f_{n,t})\,\Omega \doteq
(\psi_1,\spcpath_1)\timesIn\cdots\timesIn(\psi_n,\spcpath_n)\, 
\end{equation}
exists, and the above statements also hold. 
The limit vectors  depend (linearly and) continuously on the single 
particle states, as a consequence of the following fact: 
\begin{Lem}[Scalar product of scattering states~\cite{FGR}.] 
\label{ScattProd}
Let $\phi_k$ be single particle vectors as above with the same 
charges $\pi_k$ and spectral supports as $\psi_k$, $k=1,\ldots,n$. 
Let further $T$ be a self-intertwiner of $\pi_1\times \cdots
\times\pi_n$. Then there holds
\begin{multline} \label{eqScattProd} 
\Lsp (\phi_1,\spcpath_1)\timesIn\cdots \timesIn (\phi_n,\spcpath_n),
T  (\psi_1,\spcpath_1)\timesIn\cdots \timesIn
(\psi_n,\spcpath_n)\Rsp \\
= \chi_1\cdots \chi_n(T)\,\prod_{k=1}^n \lsp\phi_k,\psi_k\rsp,
\end{multline}
where $\chi_k$ is the standard right inverse of $\pi_k$, 
$k=1,\ldots,n$.  
\end{Lem}
Note that by the Reeh-Schlieder property~\eqref{eqReehSchlie}, every
single-particle state $\psi\in\calH_\pi$ with spectral support in some
given $V\subset \Hyp$ can be approximated by a sequence of the form
$F_\nu(f)\Omega$, where the field operators $F_\nu$ are localized in 
some fixed (arbitrary) region $\spcpath$. This allows the construction, by
continuous extension,  of scattering states 
\beq \label{eqScatt1n}
(\psi_1,\spcpath_1)\timesOut\cdots \timesOut
(\psi_n,\spcpath_n)
\eeq
for any single particle vectors $\psi_1,\ldots,\psi_n$ with compact   
mutually disjoint spectral supports and localization regions 
$\spcpath_k$ which are future-admissible for the supports of the
$\psi_k$.  
The scattering states inherit the commutation relations of the
fields, for example~\cite{FGR}   
\beq \label{eqScattStat} 
(\psi_1,\spcpath_1)\timesOut(\psi_2,\spcpath_2) = 
\Eps{\pi_1}{\pi_2}(\spcpath_1,\spcpath_2)\;
(\psi_2,\spcpath_2)\timesOut(\psi_1,\spcpath_1).
\eeq
The dependence of the scattering states on the space-time localization
regions is well-known~\cite{FGR}. But since our argument relies on an
explicit formula in the case $n=2$, we shall exhibit this formula, 
together with the proof. 
\begin{Lem}[Change of localization regions] \label{ScattRegion}
Let $\psi_1,\psi_2$ be single particle states with spectral
supports in $V_1,V_2$, respectively. Suppose that all the pairs 
$(\cck_1,\cck_2)$, $(\cck_1,\spc_2)$, and $(\spc_1,\spc_2)$ are 
future-admissible for $V_1\times V_2$. Then 
\begin{align} 
(\psi_1,\cckpath_1)\timesOut (\psi_2,\cckpath_2)&=
(\psi_1,\cckpath_1)\timesOut (\psi_2,\spcpath_2) \label{eq122} \\
&=\Eps{\pi_1}{\pi_2}(\cckpath_1,\spcpath_2)
\Eps{\pi_2}{\pi_1}(\spcpath_2,\spcpath_1)\,
(\psi_1,\spcpath_1)\timesOut (\psi_2,\spcpath_2). \label{eq112}
\end{align}
The same holds for the respective incoming scattering states if
$(\cck_1,\cck_2)$, $(\cck_1,\spc_2)$, and  
$(\spc_1,\spc_2)$ are past-admissible for $V_1\times V_2$.
\end{Lem}
\begin{Proof} 
In a first step, suppose that the vectors $\psi_i$ are at the same time of the 
form $\psi_i=F_i(f_i)\Omega$ and $\psi_i=G_i(g_i)\Omega$, 
with $F_i\in\calF_{\pi_i}(\spcpath_i)$ and
$G_i\in\calF_{\pi_i}(\cckpath_i)$, $i=1,2$. 
Then Eq.~\eqref{eq122} holds obviously.  
In order to describe the dependence on the first
localization region $\cckpath_1$, we use Eq.~\eqref{eqScattStat} to
commute the orders, then use Eq.~\eqref{eq122} to replace
$\cckpath_1$ by $\spcpath_1$, and commute back: 
\begin{align*} 
(\psi_1,\cckpath_1)\timesOut (\psi_2,\spcpath_2)&=
\Eps{\pi_1}{\pi_2}(\cckpath_1,\spcpath_2)\, 
(\psi_2,\spcpath_2)\timesOut (\psi_1,\cckpath_1) \\
&=\Eps{\pi_1}{\pi_2}(\cckpath_1,\spcpath_2)\,
(\psi_2,\spcpath_2)\timesOut (\psi_1,\spcpath_1) \\
&=\Eps{\pi_1}{\pi_2}(\cckpath_1,\spcpath_2)
\Eps{\pi_2}{\pi_1}(\spcpath_2,\spcpath_1)\,
(\psi_1,\spcpath_1)\timesOut (\psi_2,\spcpath_2).  
\end{align*}
Thus Eq.~\eqref{eq112} holds in this special case.
Now in general the $\psi_i$ will not be exactly of the form 
$\psi_i=F_i(f_i)\Omega=G_i(g_i)\Omega$, but they
can be approximated by such vectors. Then the above
equations~\eqref{eq122} and \eqref{eq112} hold by continuity.
\end{Proof}
We shall henceforth call a space-like cone $\spc_1$ {\em future- (past-) 
admissible} for a pair $V_1\times V_2$ of compact sets on the mass shell if 
the regions $\overline{\spc_1}+t\,\V(V_1)$ 
and $t\V(V_2)$ are space-like separated for large positive (negative) $t$.  
(For then there exists a cone $\spc_2$ such that $\spc_1,\spc_2$ are
admissible for $V_1\times V_2$ in the sense of the earlier
definition.)  
Since the scattering state is independent of the localization region
$\spc_2$ of $\psi_2$, we shall write $(\psi_1,\spcpath_1)\timesOut
\psi_2$ instead of $(\psi_1,\spcpath_1)\timesOut (\psi_2,\spcpath_2)$. 

We shall consider {\em special space-like cones} in the sense of Bros
and Epstein~\cite{BrosEpstein94}, referring to the reference frame 
$u$ and its rest space $\Sigma\doteq u^\perp$. 
Let $\underline{\spc}$ be an open, salient cone in 
$\Sigma$ with apex at the origin. Then its causal completion 
$\spc=\underline{\spc}''$ is a special space-like cone.  
We shall make use of the following simple observations. 
\begin{Lem} \label{admissible}

$i)$ 
Let $\spc_1,\spc_2$ be space-like cones with apices at the origin,
and let $\V_1$, $\V_2$ be compact disjoint sets in velocity space. 
If the condition "$\overline{\spc_1}+t\,\V_1$ and 
$\overline{\spc_2}+t\V_2$ are causally separated"
holds for some $t>0$ then it holds for all $t>0$. (The same is true
for negative $t$.) 

$ii)$ Let $\spc_1,\spc_2$ be special space-like cones and 
let $p_1\neq p_2\in \Hyp$. 
$\spc_1$, $\spc_2$ are future- and past- admissible for 
$(p_1,p_2)$ if, and only if, the span of $\{p_1,p_2\}$ 
has trivial intersection with the closure of the difference cone
$\spc\doteq \spc_2-\spc_1$,
\beq \label{eqpC}
\text{span}\{p_1,p_2\}\cap\overline \spc =\{0\}. 
\eeq
\end{Lem}
\begin{Proof}
Let $t,t'>0$ with $t'=\lambda t$. $\overline{\spc}_1+t\,\V_1$ and 
$\overline{\spc}_2+ t\V_2$ are causally separated if and only if 
$$
0<\big(x+t(v_2-v_1)\big)^2 \;  \equiv \;
\lambda^{-2}\,\big(\lambda x+t'(v_2-v_1)\big)^2
$$ 
for all $v_i\in \V_i$ and $x\in \spc$, where $\spc\doteq
\spc_2-\spc_1$. If this condition holds for $t$ then it also
holds for $t'$ since $x\in\spc$ iff $\lambda x\in \spc$,
$\lambda>0$. This proves $i)$. 
To prove $ii)$, let $v_i$ be the velocities in the given reference frame 
corresponding to $p_i$, namely 
\beq \label{eqvp}
v_i\doteq \big(1,\frac{\bfp_i}{\omega(\bfp_i)}\big). 
\eeq
Note that for special space-like cones, $\spc_i=(\underline{\spc_i})''$, 
$\overline{\spc_1}+tv_1$ is causally separated from $\overline{\spc_2}+tv_2$
 if and only if $\overline{\underline\spc_1}+ tv_1$ is disjoint from 
$\overline{\underline\spc_2}+ tv_2$. Thus, by part $i)$, $\spc_i$ are 
future and past-admissible for $(p_1,p_2)$ iff 
$\overline{\underline\spc_1}+ tv_1$ is disjoint from 
$\overline{\underline\spc_2}+ tv_2$ for all $t>0$ and for all $t<0$,
that is, iff 
\beq \label{eqRvC}
\RR(v_1-v_2)\cap \overline{\underline\spc}= \{0\}. 
\eeq
Since $\underline\spc$ is contained in 
$\Sigma=(1,\boldsymbol{0})^\perp$, and the span of $v_1,v_2$ is a 
time-like hyper-plane whose intersection with $\Sigma $ is just 
$\RR(v_1-v_2)$, the above equation is equivalent with 
$$
\{0\} = \text{span}\{v_1,v_2\}\cap \overline{\underline{\spc}}= 
\text{span}\{v_1,v_2\}\cap \overline{\spc}. 
$$    
But the span of $v_1,v_2$ obviously coincides with the span of
$p_1,p_2$, and the proof is complete.  
\end{Proof}
\section{Two-Particle LSZ Formulae}   \label{secLSZ} 
The LSZ-relations, which relate S-matrix elements with time-ordered 
products of fields, have been derived by Hepp within Haag-Ruelle
theory~\cite{Hepp65}, see~\cite{BuSuScatt} for a review.   
Using the arguments of~\cite{BuSuScatt}, we verify here that 
certain LSZ-relations are valid in
the case of localization in space-like cones and of braid group
statistics, under certain hypothesis on the space-time 
localization regions and momentum supports. 
Let $\pi_1$, $\pi_2\in \Delta$, let $\spc_1$ and $\spc_2$
be causally separated special space-like cones and 
$\spcpath_i$ be paths ending at $\spc_i$, and let  
\beq \label{eqEps12}
\eps\doteq  \Eps{\pi_1}{\pi_2}(\spcpath_1,\spcpath_2). 
\eeq 
For $F_1\in\calF_{\pi_1}(\spcpath_1)$ and 
$F_2\in\calF_{\pi_2}(\spcpath_2)$ let $TF_1(x)F_2(y)$ be the 
time-orderd product: 
\beq \label{eqT}
TF_1(x)F_2(y) = \theta(x^0-y^0) F_1(x)F_2(y) 
+\theta(y^0-x^0)\eps\, F_2(y)F_1(x). 
\eeq 
(Here, $\theta$ is the Heaviside function, $\theta(t)=0$ if $t < 0$
and $\theta(t)=1$ if $t\geq0$.)
Let now $V_1, V_2$ be compact subsets of the mass shell such that 
$\spc_1,\spc_2$ is both future and past-admissible for 
$V_1\times V_2$, 
let $f_1,f_2$ be test functions with momentum supports in $V_1$
and $V_2$, respectively, and denote $g_{i,t}(x)\doteq
f_{i,t}(x)-f_{i,-t}(x)$, $i=1,2$.  
Then for fixed $s>0$ and sufficiently large $t>0$ one has
\begin{align}\nonumber
& \int \d^4 x\d^4 y  \, g_{1,s}(x)\,g_{2,t}(y)\,
 T(F_1(x)F_2(y))\Omega \\ 
& \equiv 
\int \d^4 x \, g_{1,s}(x)\, \int \d^4 y \,
\big\{f_{2,t}(y)\,T(F_1(x)F_2(y)) - f_{2,-t}(y)\,T(F_1(x)F_2(y))
\big\} \Omega \nonumber \\ 
& \simeq 
\int \d^4 x \, g_{1,s}(x)\,
\int \d^4 y \, 
\big\{f_{2,t}(y)\,\eps F_2(y)F_1(x) - f_{2,-t}(y)\,F_1(x)F_2(y)
\big\} \Omega \nonumber \\ 
& = \eps \, F_2(f_{2,t}) F_1(g_{1,s})\Omega - F_1(g_{1,s})
  F_2(f_{2,t})\Omega \nonumber \\ 
& = F_1(f_{1,-s}) F_2(f_2)\Omega -F_1(f_{1,s}) F_2(f_2) \Omega.  
\label{eqLSZPosition} 
\end{align}
In the third line (``$\simeq$'' means equality up to terms which
vanish in the limit $t\to \infty$) we have used the fact that the
supports of $g_{1,s}$ and $f_{2,\pm t}$ are essentially 
contained in the regions $s\Gamma(V_1)\cup -s\Gamma(V_1)$ and 
$\pm t \Gamma(V_2)$, respectively, which are chronologically ordered as 
\begin{equation*} 
-tv_2^0 \, < \, |sv_1^0|\, 
<\,  tv_2^0 ,\quad v_i\in\Gamma(V_i),  
\end{equation*}
for fixed $s$ and sufficiently large $t>0$. 
In the last equation we have used that 
$F_1(g_{1,s})\Omega\equiv F_1(f_{1,s})\Omega-F_1(f_{1,-s})\Omega=0$
since  
$F_1(f_{1,s})\Omega$ is independent of $s$. The last expression in
Eq.~\eqref{eqLSZPosition} converges for $s\to \infty$ to 
$$
F_1(f_1)\Omega\timesIn F_2(f_2)\Omega - 
F_1(f_1)\Omega\timesOut F_2(f_2)\Omega.  
$$
On the other hand, the limit of the first line in
Eq.~\eqref{eqLSZPosition} can be calculated in Fourier space, 
noting that
\begin{equation} \label{eqgt}
\frac{\widehat f_{t}(p)-\widehat{f_{-t}}(p)}{p^2-m^2} 
 \to 2\pi i \, \delta_m(p)\, \hat f(p),\quad 
\delta_m(p)\doteq \delta(p^2-m^2)\theta(p_0),  
\end{equation}
for $t\to \infty$. 
Let $\phi$ be a state vector in $\calH_{\pi_1\times \pi_2}$ 
with compact spectral support, and let $t(x_1,x_2)$ be 
the ``time-ordered function'' 
\begin{equation} \label{eqt}
t(x_1,x_2)\doteq \Lsp \phi,T(F_1(x_1)F_2(x_2))\Omega\Rsp.   
\end{equation}
Define the amputated time-ordered function $\tamp$ by 
\begin{equation} \label{eqtamp}
\tamp(p_1,p_2)\doteq
(p_1^2-m^2)(p_2^2-m^2)\,\check t(p_1,p_2), 
\end{equation}
where $\check t$ is the inverse Fourier transform of 
$t$.\footnote{The inverse Fourier transform of a distribution 
$t$ is given by $\check{t}(\hat{f}) = t({f})$.
} 
Then the scalar product of the first line in Eq.~\eqref{eqLSZPosition} 
with $\phi$ is 
$$
t(g_{1,s}\otimes g_{2,t}) \equiv 
\tamp\big(\frac{\hat g_{1,s}}{p_1^2-m^2}\otimes \frac{\hat g_{2,t}}{p_2^2-m^2}\big)
$$
and the limit in $s,t$ is {\em formally}, by Eq.~\eqref{eqgt},  
$$
(2\pi i)^2\, (\delta_m\otimes \delta_m \cdot \tamp)(\hat f_1\otimes \hat f_2)
\equiv 
(2\pi i)^2\, \int_{\Hyp\times \Hyp}\d\mu(p_1,p_2) 
\hat f_1(p_1)\hat  f_2(p_2) \,
\tamp(p_1,p_2). 
$$
As we see in Lemma~\ref{WFT} below, this can be justified under the stated hypothesis on
$\spc_i$ and $V_i$. We then conclude from Eq.~\eqref{eqLSZPosition}:  
\begin{multline}
(2\pi i)^2\, \int_{\Hyp\times \Hyp}\d\mu(p_1,p_2) \hat f_1(p_1)\hat
  f_2(p_2) \,\tamp(p_1,p_2)|_{\Hyp\times \Hyp}
\\ 
= 
\Lsp \phi \, , \, 
F_1(f_1)\Omega\timesIn F_2(f_2)\Omega - 
F_1(f_1)\Omega\timesOut F_2(f_2)\Omega\Rsp.  \label{eqLSZ}
\end{multline}
This is the relevant LSZ relation. (See \cite[Cor.~5.10]{Araki} and
\cite{BuSuScatt} in the case of compact localization.)

It remains to show that $\tamp$ can be multiplied with the
distribution $\delta_m\otimes \delta_m$ or, equivalently, can be
restricted to the two-fold mass shell 
$$ 
\Hypp \doteq \Hyp \times \Hyp. 
$$ 
To this end it suffices to show that, over $H_m^2$, the wave front set
of $\tamp$ is disjoint from the co-normal bundle of 
$H_m^2$ (in which the wave front set of $\delta_m\otimes \delta_m$ is
contained), see~\cite[Thm.~8.2.10]{Hormander}. 
In order to determine the wave front set, we use a standard argument
involving the ``two-point'', commutator, advanced and retarded functions 
\begin{align}  
w(x_1,x_2)&\doteq   \Lsp\phi,F_1(x_1)F_2(x_2)\Omega\Rsp,
\nonumber \\ 
c(x_1,x_2)&\doteq    \Lsp\phi,F_1(x_1)F_2(x_2)-\eps F_2(x_2)F_1(x_1) \Omega\Rsp, 
\nonumber \\
a(x_1,x_2)&\doteq    \;\; \theta(x_1^0-x_2^0) \;c(x_1,x_2), 
\nonumber \\
r(x_1,x_2)&\doteq    -\theta(x_2^0-x_1^0) \; c(x_1,x_2). 
\nonumber 
\end{align}
Obviously $c=a-r$ and $t=r+w$. The amputated two-point, advanced and 
retarded functions $\wamp$, $\aamp$, $\ramp$  are defined analogously
as $\tamp$. Now the inverse Fourier transform
of $w$ is a measure in the second variable which has support in the 
energy momentum spectrum, namely
$$
\check w(\hat f_1\otimes \hat f_2) = \int \hat f_2(p)\,\d 
\Lsp\phi, F_1(f_1)E_{p}F_2\Omega\Rsp,  
$$
where $dE_p$ is the energy-momentum spectral measure. 
Hence by the mass gap condition, the amputated two-point function  
vanishes in a neighborhood of the mass shell 
(which contains no off-shell spectral points). 
Since $t\equiv r+w$, this implies that the amputated
time-ordered function $\tamp$ coincides in that neighborhood with 
the amputated retarded function $\ramp$. 
By the same token, the amputated commutator function vanishes
on a neighborhood of $\Hypp$, which implies that 
the amputated advanced and retarded functions coincide on that
neighborhood. Thus, 
\begin{equation} \label{eqTampRamp}  
\tamp = 
\aamp = \ramp \;  \text{ on a neighborhood of }  \Hypp. 
\end{equation}
Hence the wave front sets over $\Hypp$ of these distributions
coincide. Now the wave front sets of $\aamp$, $\ramp$ are related with
the supports of $a$, $r$. 
We shall use total and relative momenta $(P,p)$ as well as 
the center-of-mass and relative variables
$(X,x)$\footnote{Recall that
  $P\cdot X+p\cdot x=p_1\cdot x_1+ p_2\cdot x_2$, hence the Fourier 
transform intertwines the respective variable transformations. 
} 
\beq \label{eqPp} 
P\doteq  p_1+p_2,\quad p\doteq \frac{1}{2}\, (p_1-p_2), \quad 
X\doteq \half \,(x_1+x_2), \quad x\doteq x_1-x_2. 
\eeq 
By locality, the commutator function vanishes if $x\in \spc'$, where 
\beq \label{eqC}
\spc \doteq \spc_2 - \spc_1,  
\eeq
i.e., it has support in $(X,x)\in \RR^4\times 
\overline{\spc+({V_+}\cup {V_-})}$.  Therefore the advanced and 
retarded functions have their supports in the closed cones 
$\RR^4\times \overline{\spc+V_\pm}$, respectively. The same holds of 
course after applying the 
differential operator $(\square_1+m^2)(\square_2+m^2)$. Therefore the wave
front sets of the amputated advanced and retarded products are
contained in the same cones, 
see~\cite[Lemma~8.1.7]{Hormander} and \cite[Thm.~IX.44]{ReSi}. 
Over the neighborhood where $\tamp$ coincides  
with the amputated retarded and advanced products, the wave front set
of $\tamp$ is thus contained in the intersection of these cones, which
is 
\beq \label{eqWFt}
\RR^4\times \overline{\spc}. 
\eeq 
Now the co-normal bundle of $H_m^2$ over a point $(p_1,p_2)\in\Hypp$
is, in terms of the corresponding total and relative momentum $(P,p)$,   
\beq \label{eqCNH}
\big(T_{(P,p)}\Hypp\big)^\perp = \RR^4\times \spa\{P,p\}.  
\eeq 
It has trivial intersection with the wave front set~\eqref{eqWFt} over
$(P,p)$ if 
$\overline{\spc}$ has trivial intersection with
$\spa\{P,p\}\equiv\spa\{p_1,p_2\}$. Part $ii)$ of 
Lemma~\ref{admissible} now implies: 
\begin{Lem}[Mass shell restriction of $\tamp$.]  \label{WFT} 
If $\spc_1$ and $\spc_2$ are special space-like
cones such that $\spc_1$, $\spc_2$ is future- and past- admissible for
$V_1\times V_2\subset\Hypp$, then 
$\tamp$ can be restricted to $V_1\times V_2$. 
\end{Lem}
We have now verified that the LSZ-relation~\eqref{eqLSZ} is valid in
the case of localization in space-like cones and of braid group statistics, 
given that the localization regions
of $F_1$ and $F_2$ are causally separated and are future- and
past- admissible for the momentum supports of $f_1$ and $f_2$. 
\section{Analyticity of the Amputated Time-Ordered
  Function} \label{secAnalyt} 
The next step is to show that the amputated time-ordered function on
the left hand side of Eq.~\eqref{eqLSZ}
is the boundary value of an analytic function in the relative momentum. 
We adapt the standard arguments to verify that this holds also in the 
case of space-like cone localization and of braid group statistics. 

Let $\spcpath_i$ and let $F_i\in\calF_{\pi_i}(\spcpath_i)$ be as in
the last section, $i=1,2$, and suppose that the difference cone 
$\spc\doteq \spc_2-\spc_1$ is salient (which is the case if
$\spc_2=-\spc_1$).  
Recall from above that the corresponding advanced and retarded functions have 
their supports in the closed cones 
$(X,x)\in \RR^4\times \overline{\spc+V_\pm}$, respectively.  
It follows~\cite[Thm.~IX.16]{ReSi} that the inverse
Fourier transforms $\check a(P,p)$, $\check r(P,p)$ are, in the
$p$-variable,  boundary values  of two functions which are analytic in 
$\RR^4 + i (V_\pm\cap \spc^*)$ respectively, where
$\spc^*$ is the dual cone, namely the set of all $q\in\RR^4$ such that
$q\cdot x>0$ for all $x$ in $\overline{\spc}\setminus\{0\}$.
The same holds of course for the amputated advanced and retarded
functions. 
But these coincide by Eq.~\eqref{eqTampRamp} on a neighborhood of the
two-fold mass shell $\Hypp$. Recalling that $\Hypp$ corresponds under
the transformation~\eqref{eqPp} to the set of $(P,p)$  
such that $P^2\geq 4m^4$, $p\cdot P=0$ and 
$p^2+\frac{1}{4}P^2=m^2$, this means that $\aamp$ and
$\ramp$ coincide for 
fixed $P$ with $P^2\geq 4m^4$ on a neighborhood $U$ of the
space-like two-sphere consisting of the on-shell relative 
momenta $p$ for the given $P$,
\begin{equation} \label{eqpPhys}
M_P\doteq \{ p\in\RR^4:\; p\cdot P=0,\;p^2=m^2-\frac{1}{4}P^2\,<0\}.  
\end{equation}
The edge-of-the-wedge theorem for oblique
tubes~\cite{Epstein60,BrosIag} then asserts that the distributions 
$\aamp$ and $\ramp$ in fact are boundary values of a 
single function which is
analytic in an open set $\Theta_\spc$ (a ``local tube'') 
which contains all points $p+iq$ such that $p\in U$ and $q$ is
contained in the convex hull of $V_+\cap \spc^*$ and  $V_-\cap
\spc^*$, namely in 
\beq\label{eqCDagger}
\spc^\dagger \doteq (V_+\cap \spc^*) + (V_-\cap \spc^*), 
\eeq 
and satisfies a certain bound in norm, $\|q\|<\rho(p)$. 
Since the amputated time-ordered function $\tamp$ coincides 
with $\aamp$ and $\ramp$ on $U$, c.f.\  Eq.~\eqref{eqTampRamp}, it 
shares this analyticity property. 
Analyticity in $\Theta_\spc$ is, however, not sufficient for our later 
purpose: We shall wish to conclude from the vanishing of $\tamp$ on a 
small open set in $M_P$ that it vanishes on a larger set in $M_P$. 
We therefore 
have to consider the complexification of $M_P$. 
By the above considerations, we know that $\tamp(P,p)$ has, for 
fixed $P$, an analytic extension in $p$ into the intersection 
of $\Theta_\spc$ with the complexification of $M_P$, which we denote
by 
$$
\Theta_{P,\spc}.   
$$  
$\tamp(P,p)$ is the boundary value of this analytic function for all
$p\in M_P$ which lie at the real boundary of $\Theta_{P,\spc}$. Let 
us denote this set by
$$
M_{P,\spc}.
$$
Now as a consequence of the fact that $\spc^\dagger$ is a salient 
cone which does not contain the origin, 
$M_{P,\spc}$ does not cover all of $M_P$. (This contrasts the case 
of compact localization.) However, depending on $\spc$, it can be 
made to cover a large part of  $M_P$. 
The following proposition characterizes the the real boundary 
$M_{P,\spc}$ of the domain of analyticity of the amputated 
time-ordered function, see also~\cite[Lemma~5.1]{BrosEpstein94}. 
\begin{Prop}[Analyticity domain of $\tamp$.]  \label{Analyt}
$i)$ 
An on-shell relative momentum $p\in M_P$ 
is in $M_{P,\spc}$ if, and only if, the span of $\{P,p\}$ 
has trivial intersection with the closure of $\spc$,
\beq \label{eqPpC}
\text{span}\{P,p\}\cap\overline \spc =\{0\}. 
\eeq
$ii)$ $M_{P,\spc}$ covers $M_P$ up to two anti-podal 
neighborhoods $U_{P,\spc}\cup
-U_{P,\spc}$, where $U_{P,\spc}$ is the intersection of 
$M_P$ with the projection of $\overline {\spc}$ onto 
$P^\perp$, 
\begin{equation} \label{eqCDagP}
U_{P,\spc}= M_P\, \cap \,E_P^\perp \overline \spc. 
\end{equation}
\end{Prop}
Note that by part $ii)$ of Lemma~\ref{admissible} the 
condition~\eqref{eqPpC} is satisfied for all
$(p_1,p_2)\in V_1\times V_2$ if $\spc_1$, $\spc_2$ is future- and
past- admissible for $V_1\times V_2$.  
Note also that in $d\geq 4$ the set $M_{P,\spc}$ is connected, 
whereas in $d=3$ it has two connected components. 
\begin{Proof}
The complexification of $M_P$ consists of all $p+iq\in\CC^4$ 
with $p,q\in P^\perp$ and $p\cdot q=0$,
$p^2-q^2=m^2-\frac{1}{4}P^2$. 
The real boundary of $\Theta_{P,\spc}$ therefore consists of all
points $p\in M_P$ for which there is some 
$q\in P^\perp\cap \spc^\dagger$ orthogonal to $p$. (For then 
$$
\frac{\sqrt{\mu^2-\lambda^2q^2}}{\mu}\,p+i\lambda q 
$$ 
converges to $p$ within $\Theta_\spc$ if $\lambda\to 0$. Here, 
$\mu^2\doteq \frac{1}{4} P^2-m^2$.) In other words, $p$ belongs to 
the real boundary if, and only if, 
\beq \label{eqPiC*}\
\Pi^\perp \cap \spc^\dagger \neq \emptyset,
\eeq 
where $\Pi$ is the time-like plane spanned by $P,p$.
Since $\spc^\dagger\subset \spc^*$, it is obvious that \eqref{eqPiC*} 
implies \eqref{eqPpC}. To prove
that \eqref{eqPpC} implies \eqref{eqPiC*}, namely the ``if'' 
statement of part $i)$, we show in a  
first step that the condition~\eqref{eqPpC} allows for the 
construction of a wedge which
contains $\overline \spc\setminus\{0\}$ and whose edge has non-trivial
intersection with $\Pi$. 
The hypothesis~\eqref{eqPpC} implies that the intersection of $\Pi$ 
with the closure of the base $\underline\spc$ of our special space-like
cone is trivial. Thus, the intersection of $\Pi$ with the 
space-like hyper-surface $\Sigma\supset\underline\spc$, which is a 
one-dimensional space-like subspace $L$, has trivial 
intersection with $\overline{\underline \spc}$. 
Since $\underline \spc$ is a salient cone in the $(d-1)$-dimensional
hyper-plane $\Sigma$, there exists  
a $(d-2)$-dimensional linear hyper-plane $E$ of $\Sigma$ which 
contains the line $L$ and still has trivial  
intersection with $\overline{\underline{\spc}}$. Then $E$  divides $\Sigma$ into
two connected components and $\underline \spc$ is contained in 
one of them. 
Let $W$ be the causal completion of this half-space of $\Sigma$. 
Then $W$ is a wedge region\footnote{A {\em wedge} is a region in
  Minkowski space which arises from the standard wedge 
\begin{align*} 
W_1 &\doteq\{\,x\in\RR^d:\,|x^0|<x^1\;\}\,, 
\end{align*} 
by a Poincar\'e transformation. Its {\em edge} is the intersection of
the upper and lower bordering light-like half-planes.
}
which contains $\overline\spc\setminus \{0\}$  and whose edge, $E$,
has non-trivial intersection with $\Pi$, namely,
$\Pi\cap E  = L$. 
Our second step is to show that the existence of such a wedge implies
the claimed Eq.~\eqref{eqPiC*}. 
The upper and lower boundaries of $W$ contain two light-like 
rays $\RR l_\pm\subset \partial V_\pm$. They lie in the 
orthogonal complement of $E$ and characterize $W$ as follows: 
\beq \label{eqWll}
W=\{x\in \RR^d:\; x\cdot l_+<0,\; x\cdot l_-<0\}. 
\eeq
Now by construction, $\Pi$ and $ E$ span a $d-1$-dimensional 
time-like hyper-plane and 
$\Pi^\perp\cap E^\perp$ is a one-dimensional space-like subspace. 
Let $q$ be in the component of this line which is contained in the 
opposite wedge $-W$. Then 
$$
q=-l_+'-l_-'\quad \in\;  \Pi^\perp\cap E^\perp, 
$$  
where $l'_\pm\in\partial V_\pm$ are suitable positive multiples of 
$l_\pm$. 
The vectors $-l_\pm'$ are in
the dual of $\spc$ by Eq.~\eqref{eqWll}. Thus, the time-like vectors
$$
q_\pm \doteq -l'_\mp \pm \eps (l'_+-l'_-) \quad \in\;  V_\pm,
$$
are still contained in the dual of $\spc$ for sufficiently small 
$\eps>0$.    
We have thus found $q\in\Pi^\perp$ which can be written 
as $q=q_++q_-$ with $q_\pm$ in $V_\pm\cap \spc^*=\spc^\dagger$, i.e.,
Eq.~\eqref{eqPiC*} holds. This completes the proof of $i)$. 
Now Eq.~\eqref{eqPpC} is equivalent with 
$$ 
\RR p \, \cap \, E_P^\perp \overline\spc \;  =\;  \{0\}, 
$$
which proves $ii)$. 
\end{Proof}
\section{Proof of the Theorems} \label{secProofs} 
\subsection{The four (and higher) dimensional case.} 
\label{secProof1} 
Recalling the situation of Theorem~\ref{STrivial}, suppose there 
are sets of incoming momenta $U_1,U_2\subset \Hyp$ and of 
outgoing momenta $V_1,V_2\subset \Hyp$, all mutually disjoint, 
cf.\ Eq.~\eqref{eqMutualDisjoint}, 
and compatible with energy-momentum conservation: 
Namely, $U_1+U_2$ and $V_1+V_2$ are neighborhoods of some 
$P\in V_+$, $P^2\geq 4m^2$.  
The hypothesis is that there is no elastic two-particle scattering
from 
$U_1\times U_2$ into $V_1\times V_2$ in the channel $\pi_1\times
\pi_2\to \pi_1\times \pi_2$, namely: 
\begin{equation} \label{eqStriv'}
\Lsp \phi_1\timesIn \phi_2, \psi_1\timesOut \psi_2 \Rsp
= 0 
\end{equation} 
for all single-particle states $\phi_i$ in the sectors
$\pi_i$ with respective spectral supports in $U_i$, and all $\psi_i$
in the same sectors $\pi_i$ and with supports in $V_i$. 
Due to the assumed disjointness of the energy-momentum
supports, the state $\psi_1\timesIn \psi_2$ has 
vanishing scalar product with $\phi\In\doteq \phi_1\timesIn \phi_2$, 
and the hypothesis can be rephrased as 
\beq \label{eqInOut0} 
\Lsp \phi\In, \psi_1,\timesIn \psi_2 - \psi_1\timesOut \psi_2\Rsp  = 0  
\eeq
for all $\psi_i$ with spectral supports in $V_i$. 
We wish to show that Eq.~\eqref{eqInOut0} also holds if the spectral
supports of the $\psi_i$ are contained in other subsets 
$V_i'$ of the mass shell, with $\phi\In$ fixed.  
We shall assume that the sets $V_i$ and $V_ i'$ are small neighborhoods of 
4 on-shell momenta $p_1\neq p_2$ and $p_1'\neq p_2'$, and that 
$$
V_1'+V_2' \, \subset\,  V\doteq V_1+V_2. 
$$ 
We then choose special space-like cones $\spc_1$, $\spc_2\doteq -\spc_1$ 
such that the pair $\spc_1,\spc_2$ is future- and past- admissible 
{\em both} for $V_1\times V_2$ and for $V_1'\times V_2'$. 
\begin{Lem} 
There exists such a special space-like cone $\spc_1$. 
\end{Lem}
\begin{Proof}
Let $v_i$ and $v_i'$ be the velocities corresponding to $p_i$ and
$p_i'$, respectively, as in~Eq.~\eqref{eqvp}. 
Choose a unit vector $e$ in $\Sigma$
which is disjoint from $\RR(v_1-v_2)$ and from $\RR(v_1'-v_2')$, and let
$\underline{C}_1$ be a cone in $\Sigma$ centered at $e$, with sufficiently small
opening angle such that $\underline{C}_1$ still has trivial
intersection both with $\RR(v_1-v_2)$ and $\RR(v_1'-v_2')$. Let
$\spc_1$ be its causal completion and let $\spc_2\doteq -\spc_1$. Then
$\spc_1, \spc_2$ are future- and past- 
admissible for $\{(v_1,v_2)\}$ and for $\{(v_1',v_2')\}$, see
Eq.~\eqref{eqRvC} in the proof of Lemma~\ref{admissible}. The same
holds for the set $V_1\times V_2$ and $V_1'\times V_2'$ if these are
small enough.
\end{Proof}
The difference cone $\spc\doteq  \spc_2-\spc_1$ is now just 
$\spc \equiv -\spc_1$.     
Let us denote by $M_{V,\spc}$ the set of pairs of on-shell 
momenta whose total momentum is contained 
 in $V$ and whose relative momenta are contained in the real
boundary of the analyticity domain of the corresponding amputated time-ordered
functions. By virtue of Proposition~\ref{Analyt} this set is given by 
\beq \label{eqMVC}
M_{V,\spc}\doteq \big\{ (p_1',p_2')\in\Hyp\times \Hyp:\; 
p_1'+p_2'\in V,\, \spa\{p_1',p_2'\}\cap \overline{\spc}=\{0\}\big\}. 
\eeq 
Now Eq.~\eqref{eqInOut0} holds for all $\psi_i$ of the form
$\psi_i=F_i(f_i)\Omega$, where $F_i\in\calF_{\pi_i}(\spc_i)$ and 
where $f_i$ are test functions with respective momentum supports in
$V_i$.  
The LSZ relation~\eqref{eqLSZ} then implies that 
\beq \label{eqTamp0}
\tamp(p_1,p_2)=0
\eeq 
for all $p_i\in V_i$. 
Here, $\tamp$ is the amputated time-ordered function, with
$t(x_1,x_2)$ being defined as in Eq.~\eqref{eqt} with 
$\phi=\phi\In$ and $F_i$ as above. 
Since our pair of cones $\spc_1,\spc_2$ is future- and past- admissible for
$V_1\times V_2$,
the remark after Proposition~\ref{Analyt} implies that the 
relative momenta corresponding to $V_1\times V_2$ are, for all 
$P'\in V$, 
contained in the boundary $M_{P',\spc}$ of the domain of analyticity of 
$\tamp(P',\cdot)$. 
Now in four- (and higher) dimensional space-time, $M_{P',\spc}$ is 
connected and therefore Eq.~\eqref{eqTamp0} and 
analyticity of $\tamp$ imply that $\tamp(P',p')$ vanishes for all 
$P'\in V$ and $p'\in M_{P',\spc}$,  
that is to say, $\tamp$ vanishes on $M_{V,\spc}$. 
Since our cones $\spc_1,\spc_2$ are future- and past- admissible 
also for the set $V_1'\times V_2'$, the 
latter is also contained in the set $M_{V,\spc}$
where we have just shown that $\tamp$ vanishes. 
Again by the LSZ relation~\eqref{eqLSZ}, we conclude that 
Eq.~\eqref{eqInOut0} holds for all 
$\psi_i$ of the form $\psi_i=F_i(f_i)\Omega$, where
$F_i\in\calF_{\pi_i}(\spc_i)$ and 
the momentum supports of $\hat f_i$ are contained in $V_i'$.  
By continuity of the scattering states and the Reeh-Schlieder
property, we arrive at the following result: 
\begin{Prop}[Triviality of the S-matrix element in $d\geq 4$.] 
\label{PhiPsi1Psi20} 
Let $V_i$ and $U_i$ be as in Theorem~\ref{STrivial}, $i=1,2$, and let 
$V_1',V_2'$ be disjoint, sufficiently small, subsets of the 
mass shell such that  $V_1'+V_2'\subset V=V_1+V_2$. 
Then there holds 
\beq \label{eqInOut0'} 
\Lsp \phi_1\timesIn\phi_2, \psi_1\timesIn \psi_2\Rsp  =  
\Lsp  \phi_1\timesIn\phi_2, \psi_1\timesOut \psi_2\Rsp  
\eeq
for all single-particle states $\psi_i$
with spectral supports in $V_i'$ and for all $\phi_i$
with spectral supports contained in the sets $U_i$, $i=1,2$. 
\end{Prop}
By Lorentz covariance, the same holds if $U_1$ and $U_2$ are subject
to a common Lorentz transformation. 
Since $V$ was arbitrary in Theorem~\ref{STrivial},
it follows by linearity and continuity that
Eq.~\eqref{eqInOut0'} holds for all single-particle states $\phi_i$,
$\psi_i$.  
In other words, we have shown that $E^{(2)}\In
SE^{(2)}\In=E^{(2)}\In$, where $E^{(2)}\In$ is the projection
onto the closed subspace of incoming two-particle scattering states,
and $S$ is the S-matrix,  
$$
S:\; \psi_1\timesIn \psi_2 \; \mapsto \; \psi_1\timesOut \psi_2.
$$
But then the assumption that $SE^{(2)}\In$
has other components than in $E^{(2)}\In\calH$ contradicts the isometry of the
S-matrix. It then follows that $SE^{(2)}\In=E^{(2)}\In$, i.e.\ 
Eq.~\eqref{eqIn=Out} holds. This concludes the proof of
Theorem~\ref{STrivial}. 
\subsection{The three-dimensional case.} \label{secProof2}
In three space-time dimensions two complications arise: Firstly,
the scattering states depend on the space-time localization 
regions, and secondly, the set of analyticity $M_{V,\spc}$ has  
two connected components and the above argument only leads to the 
conclusion that $\tamp$ vanishes on the component which contains 
$V_1\times V_2$.  
 
The hypothesis of Theorem~\ref{WW} is: 
There are sets of incoming momenta $U_1,U_2\subset \Hyp$ and of 
outgoing momenta $V_1,V_2\subset \Hyp$, all mutually disjoint, 
cf.\ Eq.~\eqref{eqMutualDisjoint}, 
and compatible with energy-momentum conservation, 
c.f.\ Eq.~\eqref{eqMutualDisjoint}, and there is {\em some} path 
$\spcpath_0$, future-admissible for $V_1\times V_2$,  such that 
\begin{equation} \label{eqStrivBn}
\Lsp \phi_1\timesIn \phi_2, (\psi_1,\spcpath_0)\timesOut \psi_2 \Rsp
= 0 
\end{equation} 
holds for all single-particle states $\phi_i$ in the 
sectors $\pi_i$ with respective spectral supports in 
$U_i$,\footnote{The
space-time localization regions for the state 
$\phi_1\timesIn\phi_2$ will not play any significant role.
}  
and all $\psi_i$ in the same sectors $\pi_i$ and with supports 
in $V_i$, $i=1,2$. 
We may assume that the spectral supports $U_i$ are small 
enough as to satisfy 
\beq \label{eqUU}
U_1+U_2\subset V\equiv V_1+V_2.
\eeq

Suppose that $\spcpath_1$ is a path ending at
a special space-like cone $\spc_1$ such that
\\  
{\bf (a)} $\spc_1$ is future- and past- admissible both for $V_1\times V_2$ and for 
$U_1\times U_2$. (Then $V_1\times V_2$ and $U_2\times U_2$ are contained 
in $M_{V,\spc}$, $\spc\doteq -\spc_1$, by the remark after Prop.~\ref{Analyt}.)
\\
{\bf (b)} $V_1\times V_2$ and $U_1\times U_2$ are in one and 
the same connected component of $M_{V,\spc}$. 
\\
{\bf (c)} The path $\spcpath_1$ is equivalent with $\spcpath_0$ for
outgoing scattering 
states with momentum supports in $V_1\times V_2$, namely,  
\beq\label{eqCC1} 
(\psi_1,\spcpath_1)\timesOut \psi_2 = (\psi_1,\spcpath_0)\timesOut
\psi_2 
\eeq
holds for all $\psi_i$ with spectral supports in $V_i$, $i=1,2$. 

Due to (c), we may substitute $\spcpath_1$ for $\spcpath_0$ 
right at the beginning of our discussion, namely in 
Eq.~\eqref{eqStrivBn}, and as in the last section we conclude that 
\beq \label{eqInOut0Bn} 
\Lsp \phi\In, (\psi_1,\spcpath_1)\timesIn \psi_2 - 
(\psi_1,\spcpath_1)\timesOut \psi_2\Rsp  = 0  
\eeq
holds for all $\psi_i$ of the form $\psi_i=F_i(f_i)\Omega$, where
$F_i\in\calF_{\pi_i}(\spcpath_i)$ with $\spc_2\doteq -\spc_1$ and
$\spcpath_2$ arbitrary, and
where the momentum supports of $f_i$ are contained in $U_i$.  
Continuity and the Reeh-Schlieder property then imply: 
\begin{Prop}[Local triviality of the S-matrix element in $d=3$.] 
\label{PhiPsi1Psi20'} 
Suppose the spectral supports $U_i$ are small enough as to 
satisfy $U_1+U_2\subset V$.
Then Eq.~\eqref{eqInOut0Bn} holds for all single-particle states $\psi_i$
with momentum supports in $U_i$, $i=1,2$, and for all paths
$\spcpath_1$ with the above properties (a), (b), (c).  
\end{Prop}
We now show that this result is incompatible with braid 
group statistics, the idea being as follows. 
We construct two paths $\spcpath_1$ and $\cckpath_1$ which both
satisfy the properties (a) through (c), and such that for all 
$\psi_i$ with spectral supports in $U_i$ the outgoing scattering 
states are invariant under a change of localization regions from 
$\spcpath_1$ to $\cckpath_1$, whereas 
the corresponding incoming scattering states differ by a monodromy
operator $\eps_M$, that is 
\begin{equation} 
(\psi_1,\spcpath_1)\timesOut \psi_2= 
(\psi_1,\cckpath_1)\timesOut\psi_2, \quad 
(\psi_1,\spcpath_1)\timesIn\psi_2 = 
\eps_M \; (\psi_1,\cckpath_1)\timesIn\psi_2. 
\end{equation}
Together with the relations asserted by
Proposition~\ref{PhiPsi1Psi20'}, namely  
\begin{align} 
\Lsp \phi\In, (\psi_1,\spcpath_1)\timesIn \psi_2\Rsp &= 
\Lsp \phi\In, (\psi_1,\spcpath_1)\timesOut\psi_2\Rsp\quad
\text{ and}\\ 
 \Lsp \phi\In, (\psi_1,\cckpath_1)\timesIn\psi_2\Rsp &= 
\Lsp \phi\In, (\psi_1,\cckpath_1)\timesOut\psi_2\Rsp , 
\end{align}
it follows that 
\beq \label{eq1-eps} 
  \lsp \phi\In, (\unity-\eps_M)\, (\psi_1,\cckpath_1)\timesIn\psi_2\Rsp=0. 
\eeq
This is in conflict with braid group statistics.  

In a first step, we show that there exists a path $\spcpath_1$ with
the properties (a) through (c). 
Consider the set $J$ of normalized velocity differences contained in
$V_1\times V_2$,
$$
J\doteq \Big\{\frac{v(p_2)-v(p_1)}{\|v(p_2)-v(p_1)\|}\,:\; p_i\in V_i\Big\},
$$
where $v(p)\doteq (1,\bfp/\omega(\bfp))$ is the velocity 
corresponding to $p$. Similarly, let $I$ be the set of normalized 
velocity differences contained in $U_1\times U_2$. $I$ and $J$ are
closed intervals in the unit circle of the rest space $\Sigma$, 
$$
S^1\doteq \Sigma\cap H.
$$
If the sets $U_i$ and $V_i$ are sufficiently small and $V\equiv
V_1+V_2$ is a small neighborhood of some $P\in V_+$, then $U_i$ and
$V_i$ are disjoint neighborhoods of points on the circle $M_P$ 
in $P^\perp$. 
Therefore $I$, $-I$, $J$ and $-J$ are all disjoint.  
By Lemma~\ref{C}, we may assume that $\spc_0$ is a {\em special} space-like
cone, $\spc_0=\underline{\spc}_0''$, with small opening angle, 
centered at some ray $\RR^+e_0$, $e_0\in S^1$. 
The path $\spcpath_0$ then corresponds, with the
identification~\eqref{eqRH}, to a path $\tilde{e}_0$ in $S^1$ ending
at $e_0$. Now $\spc_0$ is future-admissible for $V_1\times V_2$ (as 
assumed) if and only if $e_0$ is in $S^1\setminus J$, see Eq.~\eqref{eqCI}. 
We wish to construct a special space-like cone 
$\spc_1$, centered at some unit vector $e\in S^1$, satisfying 
properties (a) through (c).  
Again by Eq.~\eqref{eqCI}, $\spc_1$ satisfies property (a) iff 
$e$ is in one of the four connected components of 
$S^1\setminus \big\{I\cup -I\cup J \cup -J\big\}$, and property 
(b) iff $e$ is contained in the open interval either between 
$I$ and $-J$ or between $-I$ and $J$. 
To be specific, we choose $e$ in the interval between $I$ and
$-J$. 
Define now the path $\tilde{e}$ corresponding to $\spcpath_1$ by
appending to $\tilde{e}_0$ a path from $e_0$ to $e$ which is 
contained in $S^1\setminus J$. Then according to Lemma~\ref{C}, 
$\spcpath_1$ satisfies property (c), too. 

\begin{figure}[ht] 
 \label{Fig2}
\psfrag{C}{$\spcpath_1$}
\psfrag{K}{$\cckpath_1$}
\psfrag{I}{$I$}
\psfrag{J}{$J$}
\psfrag{-I}{$-I$}
\psfrag{-J}{$-J$}
\psfrag{g}{$\gamma$}
\psfrag{e}{$e$}
\psfrag{-e}{$-e$}
\begin{center}
\epsfxsize15ex 
\epsfbox{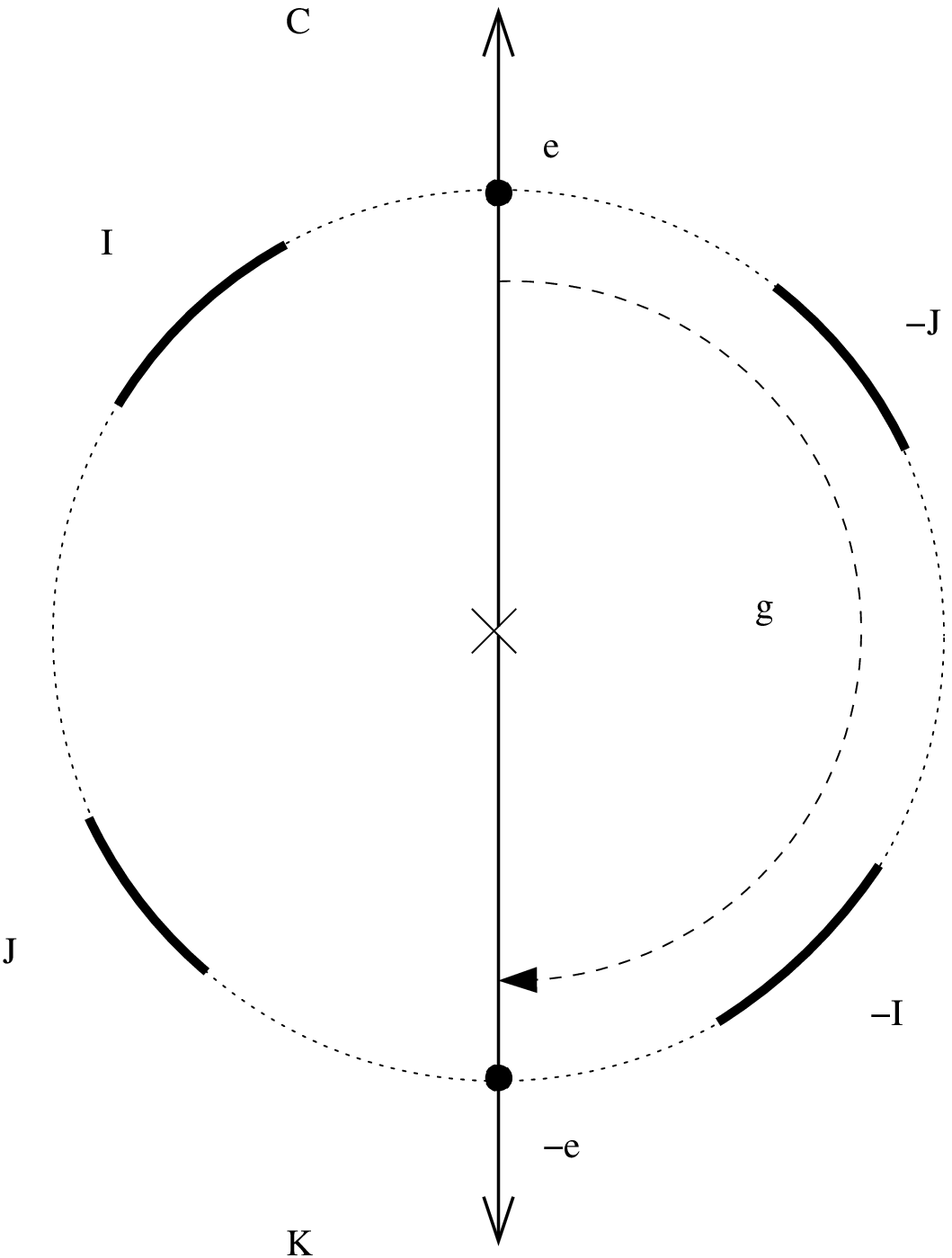}
\caption{The unit circle in the hyper-plane $\Sigma$. $\spc_1$ is
  centered at $e$ and $\cck_1$ is centered at $-e$. $\spcpath_1$ and
  $\cckpath_1$ differ by the path $\gamma$.}
\end{center}
\end{figure}
We now define a second path $\cckpath_1$ as follows. Let $\cck_1$ be
the special space-like cone $\cck_1\doteq-\spc_1$. It is 
centered at $-e$, 
located in the open interval between $-I$ and $J$, and
also satisfies properties (a) and (b).  
Define the path corresponding to $\cckpath_1$ by
appending to $\tilde{e}$ a path $\gamma$ from $e$ to $-e$ which
neither crosses $I$ nor $J$, see Figure~2. (This is possible since $I$ and $J$ are
in the same connected component of $S^1\setminus\{e,-e\}$.) 
Then according to Lemma~\ref{C}, 
\beq \label{eqKCout}
(\psi_1,\spcpath_1)\timesOut\psi_2=
(\psi_1,\cckpath_1)\timesOut\psi_2 
\eeq
holds for all $\psi_i$ with spectral supports in $V_i$, as well as 
for all $\psi_i$ with spectral supports in $U_i$, $i=1,2$. 
(See Figure~4 in the appendix.) 
In particular, $\cckpath_1$ satisfies property (c), and therefore
Prop.~\ref{PhiPsi1Psi20'} applies. 
But the path $\gamma$, which connects $\spcpath_1$ and
$\cckpath_1$, must inevitably cross the interval $-I$. This implies
that for $\psi_i$ with spectral supports in $U_i$ the incoming 
scattering states 
$(\psi_1,\cckpath_1)\timesIn\psi_2$
and $(\psi_1,\spcpath_1)\timesIn\psi_2$ differ by some monodromy
operator.  
To calculate this operator let us assume, to be specific, that the 
intervals $I,J,-I,-J$ are neighbors in the mathematically
positive sense, as in Figure~2. 
Then our path from $\spc_1$ (centered
at $e$) to $\cck_1$ (centered a $-e$) must be such that  
$\theta(\cckpath_1)=\theta(\spcpath_1)-\pi$ in order not to cross 
$I$ or $J$. 
\begin{figure}[ht] 
 \label{Fig3}
\psfrag{C1}{$\spcpath_1$}
\psfrag{I1}{$\cckpath_1$}
\psfrag{K2}{$\cckpath_2$}
\psfrag{U1}{$-t\Gamma_1$}
\psfrag{U2}{$-t\Gamma_2$}
\begin{center}
\epsfxsize60ex 
\epsfbox{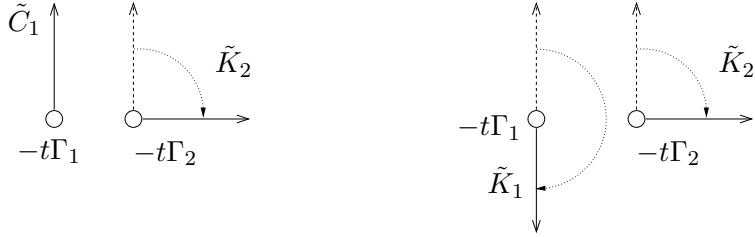}
\caption{The space-like hyper-plane $\Sigma-tu$,
  $t>0$, with $\Gamma_i\doteq \Gamma(U_i)$. 
$(\spc_1,\cck_2)$ and $(\cck_1,\cck_2)$ are past-admissible for $U_1\times U_2$. } 
\end{center}
\end{figure}
To calculate the behavior of the {incoming} scattering state under 
this change, choose a path $\cckpath_2$ such that $(\spc_1,\cck_2)$
and $(\cck_1,\cck_2)$ are past-admissible for $U_1\times U_2$ and such
that  
$$
\theta(\cckpath_1)\, < \,
\theta(\cckpath_2) \, < \, \theta(\spcpath_1) \equiv
\theta(\cckpath_1)+\pi ,   
$$ 
see Figure~3. Then 
$\Eps{\pi_1}{\pi_2}(\spcpath_1,\cckpath_2)=\Eps{\pi_1}{\pi_2}$
and $\Eps{\pi_2}{\pi_1}(\cckpath_2,\cckpath_1)=\Eps{\pi_2}{\pi_1}$,
hence 
\begin{align*}
(\psi_1,\spcpath_1)\timesIn\psi_2 & \equiv 
(\psi_1,\spcpath_1)\timesIn(\psi_2,\cckpath_2)\\
&= 
\Eps{\pi_1}{\pi_2}(\spcpath_1,\cckpath_2)\,
\Eps{\pi_2}{\pi_1}(\cckpath_2,\cckpath_1)\; 
(\psi_1,\cckpath_1)\timesIn(\psi_2,\cckpath_2) \\
&= 
\Eps{\pi_1}{\pi_2}\,\Eps{\pi_2}{\pi_1}\; 
(\psi_1,\cckpath_1)\timesIn\psi_2. 
\end{align*}
That is, the incoming scattering state in fact picks up the 
monodromy operator
\beq \label{eqMonodromy}
\eps_M \doteq 
\Eps{\pi_1}{\pi_2}\,\Eps{\pi_2}{\pi_1}
\eeq
under a change from $\spcpath_1$ to $\cckpath_1$. So
Eq.~\eqref{eq1-eps} in fact holds, with $\eps_M$ as above. 

We now show that this is in conflict with braid group statistics.  
First consider $\pi_1=\pi_2=\pi$, where $\pi$ is an Anyonic sector. Then
the monodromy operator equals $\omega_\pi^2$ times the unit operator, and 
Eq.~\eqref{eq1-eps} implies $\omega_\pi^2=1$, since $ \Lsp \phi\In, 
(\psi_1,\cckpath_1)\timesIn\psi_2\Rsp\neq 0$ for suitable
$\phi\In$. This proves Theorem~\ref{WW} in the Abelian case. 
In the non-Abelian case, where $\eps_M$ acts non-trivially, 
recall that the sector $\pi_1\times \pi_2$ contains a finite
number of irreducible sectors, each with finite
multiplicity~\cite{FRSII}. For every sub-representation 
$\sigma$ contained in $\pi_1\times \pi_2$, let $E_\sigma$ be the 
corresponding projection.\footnote{Choose an orthonormal basis 
$T_{\sigma,i}$, $i=1,\ldots,N_\sigma$, in the Hilbert space of intertwiners 
from $\sigma$ to $\pi_1\times \pi_2$; Then 
$E_\sigma\doteq \sum_{i=1}^{N_\sigma}  T_{\sigma,i}T_{\sigma,i}^*$.
} 
By Lemma 3.3 in \cite{FRSII}, these projections diagonalize the
monodromy operator: 
$$
E_\sigma \, \eps_M =
\frac{\omega_\sigma}{\omega_1\omega_2}\, E_\sigma, 
$$
where we have written $\omega_i\doteq \omega_{\pi_i}$. 
Now $\sum_\sigma E_\sigma$, where the finite sum goes over all
irreducible sub-representations of $\pi_1\times \pi_2$, is the
unit operator in $\calH_{\pi_1\times \pi_2}$ (and also is a 
self-intertwiner for $\pi_1\times \pi_2$). Denoting $\psi\In\doteq
(\psi_1,\cckpath_1)\timesIn\psi_2$, Eq.~\eqref{eq1-eps} then implies 
\begin{align*}
0 = 
\sum_\sigma \lsp \phi\In,E_\sigma \, (\unity-\eps_M)\, \psi\In \rsp 
= \sum_\sigma \big(1-\frac{\omega_\sigma}{\omega_1\omega_2}\big) 
\,  \chi(E_\sigma)\lsp \phi\In,\psi\In \rsp.  
\end{align*}
We have used Lemma~\ref{ScattProd} and have written $\chi\doteq
\chi_1\chi_2$. Now $\psi\In$ is certainly not orthogonal to all allowed
$\phi\In$, and therefore the above equation yields
$$
\sum_\sigma \chi(E_\sigma) = 
\sum_\sigma \frac{\omega_\sigma}{\omega_1\omega_2}\; 
\chi(E_\sigma) . 
$$
But $\chi(E_\sigma)$ is positive, while
$\frac{\omega_\sigma}{\omega_1\omega_2}$ has modulus one. 
Therefore all factors $\frac{\omega_\sigma}{\omega_1\omega_2}$ must equal
one, and we conclude: 
\begin{Prop}
For every irreducible sub-representation $\sigma\in\Delta$ of 
$\pi_1\times \pi_2$ there holds 
$$
\omega_\sigma=\omega_{\pi_1}\omega_{\pi_2}.
$$
\end{Prop}
Let us consider the case $\pi_1=\pi$ and $\pi_2=\bar\pi$, and recall
that the representation $\pi\times \bar \pi$ contains the vacuum 
representation
$\pi_0$, which has statistics parameter 1. Since $\omega_{\bar\pi}$
and $\omega_\pi$ coincide~\cite{FM2}, we conclude that $\omega_\pi^2=1$. 
This concludes the proof of
Theorem~\ref{WW}. 
\paragraph{Acknowledgments.}
It is a pleasure for J.M.\  to thank Detlev Buchholz for having 
taught him, among other
things, to be sceptical against any ad-hoc assumptions (concerning in
particular the S-matrix for Plektons). 
Further, J.M.\ gratefully acknowledges financial support by the 
Brazilian Research Council CNPq.   
\appendix
\setcounter{Thm}{0}
\setcounter{equation}{0}
\renewcommand{\theequation}{\thesection.\arabic{equation}}
\renewcommand{\theThm}{\thesection.\arabic{Thm}}
\section{Local independence on the localization regions of scattering
  states.}  
{}For completeness sake, we verify here the well-known fact that the
scattering states are {\em locally} independent of the space-time
localization regions $\spcpath$ in three-dimensional space-time. 
Let $V_1$, $V_2$ be compact disjoint subsets of the mass shell, and
let $J$ be the set of normalized velocity differences contained in
$V_1\times V_2$,
$$
J\doteq \Big\{\frac{v(p_2)-v(p_1)}{\|v(p_2)-v(p_1)\|}\,:\; p_i\in V_i\Big\},
$$ 
where $v(p)\doteq (1,\bfp/\omega(\bfp))$ is the velocity 
corresponding to $p$. Assuming that $V_i$ are sufficiently small, 
$J$ is a closed, simply connected, interval in the intersection of
the space-like directions $\Spd$ with the rest space $\Sigma$, 
$$
S^1\doteq \Sigma\cap H.
$$ 
Let now $\spc_1$ be a space-like cone with apex at the origin which 
is future-admissible for $V_1\times V_2$. By part $i)$ of 
Lemma~\ref{admissible}, this is the case if, and only if, the 
closure of its set of space-like 
directions $\spc_1^\Spd \equiv \spc_1 \cap \Spd$ is contained in the
causal complement of $J$ in $\Spd$,\footnote{Note that 
if $\spc_1$ is a special space-like cone,
$\spc_1=(\underline{\spc}_1)''$ with $\underline{C}_1\subset \Sigma$,
then this condition is equivalent to 
\beq \label{eqCI}
\overline{\underline{\spc}_1}\cap J=\emptyset. 
\eeq
} 
which we shall denote by
$J'$ and which is also simply connected.
Recall from Eq.~\eqref{eqRH} and thereafter that we have 
identified a path $\spcpath_1$ of space-like cones ending at $\spc_1$  
with a sheet in the universal covering space of $\Spd$ 
over $\spc_1^H\cong \spc$, that is,
with a set of (homotopy classes of) paths in $\Spd$ ending at $\spc_1^H$.
\begin{Lem} \label{C} 
Let $\spcpath_1$ and $\cckpath_1$ be paths ending at respective space-like cones
$\spc_1$ and $\cck_1$ which have their apices at the origin and
are future-admissible for $V_1\times V_2$. If the sheets
$\spcpath_1^H$ and $\cckpath_1^H$ differ by a path in $J'$, then for 
all $\psi_i$ with spectral supports in $V_i$ there holds 
\beq \label{eqOutCK}
(\psi_1,\spcpath_1)\timesOut\psi_2=(\psi_1,\cckpath_1)\timesOut\psi_2. 
\eeq 
\end{Lem}
\begin{Proof} 
We first define a localization region $\spc_2$ for the state $\psi_2$,
namely: Let $\spc_2$ be a space-like cone with apex at the origin
such that $\spc_2^H$ is contained in the causal completion of
$J$. Then the pairs $\spc_1,\spc_2$ and $\cck_1,\spc_2$ both are
future-admissible for $V_1\times V_2$. Now for any path $\spcpath_2$ ending at $\spc_2$ the 
relative winding number of $\spcpath_2$ with respect to $\spcpath_1$
coincides with that of $\spcpath_2$ w.r.t.\ $\cckpath_1$, i.e.,
$N(\spcpath_1,\spcpath_2)= N(\cckpath_1,\spcpath_2)$. Take for example
$\spcpath_2$ such that this winding number is $-1$, see Figure~4. 
\begin{figure}[ht] 
 \label{Fig4}
\psfrag{C1}{$\spcpath_1$}
\psfrag{J2}{$\spcpath_2$}
\psfrag{I1}{$\cckpath_1$}
\psfrag{U1}{$t\Gamma_1$}
\psfrag{U2}{$t\Gamma_2$}
\begin{center}
\epsfxsize60ex 
\epsfbox{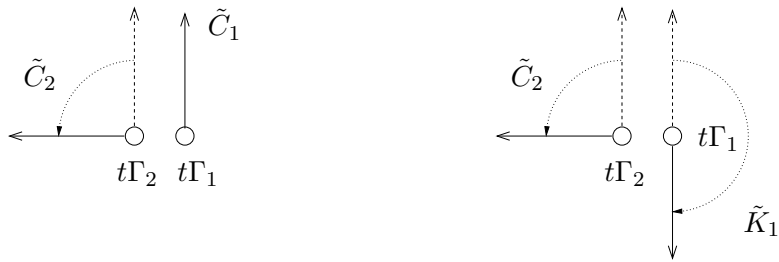}
\caption{The space-like hyper-plane $\Sigma+tu$,
  $t>0$, with $\Gamma_i\doteq \Gamma(U_i)$. 
$(\spc_1,\spc_2)$ and $(\cck_1,\spc_2)$ are 
future-admissible for $U_1\times U_2$.}
\end{center}
\end{figure}
Then
$N(\spcpath_2,\cckpath_1)=0$ and therefore by Eq.~\eqref{eqEpsN} 
$\Eps{\pi_1}{\pi_2}(\spcpath_1,\spcpath_2)=\Eps{\pi_2}{\pi_1}^{-1}$
and
$\Eps{\pi_2}{\pi_1}(\spcpath_2,\cckpath_1)=\Eps{\pi_2}{\pi_1}$. Hence 
Lemma~\ref{ScattRegion} implies that for $\psi_i\in\calH_{\pi_i}$
there holds 
\begin{align*}
(\psi_1,\spcpath_1)\timesOut\psi_2 & \equiv 
(\psi_1,\spcpath_1)\timesOut(\psi_2,\spcpath_2)= 
\Eps{\pi_1}{\pi_2}(\spcpath_1,\spcpath_2)\,
\Eps{\pi_2}{\pi_1}(\spcpath_2,\cckpath_1)\; 
(\psi_1,\cckpath_1)\timesOut(\psi_2,\spcpath_2) \\
&= (\psi_1,\cckpath_1)\timesOut(\psi_2,\spcpath_2) 
 \equiv (\psi_1,\cckpath_1)\timesOut\psi_2.  
\end{align*}
That is, the outgoing scattering state is in fact invariant under a
change of localization region from $\spcpath_1$ to $\cckpath_1$, as
claimed.
\end{Proof}
\providecommand{\bysame}{\leavevmode\hbox to3em{\hrulefill}\thinspace}
\providecommand{\MR}{\relax\ifhmode\unskip\space\fi MR }
\providecommand{\MRhref}[2]{%
  \href{http://www.ams.org/mathscinet-getitem?mr=#1}{#2}
}
\providecommand{\href}[2]{#2}

\end{document}